\documentclass{emulateapj}

\usepackage{color}
\usepackage{amsmath,natbib}
\usepackage{graphicx,epsfig}
\input{epsf}
\usepackage{epstopdf}
\usepackage{longtable,lscape}
\usepackage{multirow}
\usepackage{appendix}
\DeclareGraphicsExtensions{.jpg,.pdf,.png,.eps,.ps}
\newcommand{\sigoneam}{\ensuremath{\Sigma_{1'}}}
\newcommand{\spitzer}{{\it Spitzer}}
\newcommand{\wise}{{\it WISE}}
\newcommand{\degs}{\ensuremath{\mathrm{deg}^2}}
\newcommand{\zphot}{\ensuremath{z_\mathrm{phot}}}
\newcommand{\zspec}{\ensuremath{z_\mathrm{spec}}}
\newcommand{\zcomb}{\ensuremath{z_\mathrm{comb}}}
\newcommand{\zmed}{\ensuremath{z_\mathrm{med}}}
\newcommand{\pblank}{\ensuremath{P_\mathrm{blank}}}
\newcommand{\sigstat}{\ensuremath{\sigma_\mathrm{stat}}}
\newcommand{\sigsys}{\ensuremath{\sigma_\mathrm{sys}}}

\newcommand{\ndof}{\ensuremath{N_\mathrm{dof}}}
\newcommand{\chired}{\ensuremath{\chi^2_\mathrm{red}}}
\newcommand{\planck}{{\sl Planck}}

\begin{document}
\title{Redshifts, Sample Purity, and BCG Positions for the Galaxy Cluster Catalog from the  first 720 Square Degrees of the South Pole Telescope Survey}
 
\altaffiltext{\Michigan}{Department of Physics, University of Michigan, 450 Church Street, Ann  
Arbor, MI, 48109}
\altaffiltext{\Munich}{Department of Physics,
Ludwig-Maximilians-Universit\"{a}t,
Scheinerstr.\ 1, 81679 M\"{u}nchen, Germany}
\altaffiltext{\ExcellenceCluster}{Excellence Cluster Universe,
Boltzmannstr.\ 2, 85748 Garching, Germany}
\altaffiltext{\CfA}{Harvard-Smithsonian Center for Astrophysics,
60 Garden Street, Cambridge, MA 02138}
\altaffiltext{\KICPChicago}{Kavli Institute for Cosmological Physics,
University of Chicago,
5640 South Ellis Avenue, Chicago, IL 60637}
\altaffiltext{\PhysicsUChicago}{Department of Physics,
University of Chicago,
5640 South Ellis Avenue, Chicago, IL 60637}
\altaffiltext{\UChicago}{University of Chicago,
5640 South Ellis Avenue, Chicago, IL 60637}
\altaffiltext{\UPenn}{Department of Physics and Astronomy,University of Pennsylvania, Philadelphia, PA 19104}
\altaffiltext{\Harvard}{Department of Physics, Harvard University, 17 Oxford Street, Cambridge, MA 02138}
\altaffiltext{\EFIChicago}{Enrico Fermi Institute,
University of Chicago,
5640 South Ellis Avenue, Chicago, IL 60637}
\altaffiltext{\IAPFrance}{Institut d'Astrophysique de Paris, UMR 7095
CNRS, Universit\'e Pierre et Marie Curie, 98 bis boulevard Arago, F-75014
Paris, France}
\altaffiltext{\Miss}{Department of Physics and Astronomy, University of Missouri, 5110 Rockhill Road, Kansas City, MO 64110}
\altaffiltext{\AAUChicago}{Department of Astronomy and Astrophysics,
University of Chicago,
5640 South Ellis Avenue, Chicago, IL 60637}
\altaffiltext{\ANL}{Argonne National Laboratory, 9700 S. Cass Avenue, Argonne, IL, USA 60439}
\altaffiltext{\NIST}{NIST Quantum Devices Group, 325 Broadway Mailcode 817.03, Boulder, CO, USA 80305}
\altaffiltext{\PUC}{Departamento de Astronomia y Astrosifica, Pontificia Universidad Catolica,
Chile}
\altaffiltext{\McGill}{Department of Physics,
McGill University,
3600 Rue University, Montreal, Quebec H3A 2T8, Canada}
\altaffiltext{\Berkeley}{Department of Physics,
University of California, Berkeley, CA 94720}
\altaffiltext{\UFlorida}{Department of Astronomy, University of Florida, Gainesville, FL 32611}
\altaffiltext{\Colorado}{Department of Astrophysical and Planetary Sciences and Department of Physics,
University of Colorado,
Boulder, CO 80309}
\altaffiltext{\NASA}{Department of Space Science, VP62,
NASA Marshall Space Flight Center,
Huntsville, AL 35812}
\altaffiltext{\Davis}{Department of Physics, 
University of California, One Shields Avenue, Davis, CA 95616}
\altaffiltext{\LBNL}{Physics Division,
Lawrence Berkeley National Laboratory,
Berkeley, CA 94720}
\altaffiltext{\Caltech}{California Institute of Technology, 1200 E. California Blvd., Pasadena, CA 91125}
\altaffiltext{\Arizona}{Steward Observatory, University of Arizona, 933 North Cherry Avenue, Tucson, AZ 85721}
\altaffiltext{\MIT}{Kavli Institute for Astrophysics and Space
Research, Massachusetts Institute of Technology, 77 Massachusetts Avenue,
Cambridge, MA 02139}
\altaffiltext{\MPE}{Max-Planck-Institut f\"{u}r extraterrestrische Physik,
Giessenbachstr.\ 85748 Garching, Germany}
\altaffiltext{\CaseWestern}{Physics Department, Center for Education and Research in Cosmology 
and Astrophysics, 
Case Western Reserve University,
Cleveland, OH 44106}
\altaffiltext{\Minnesota}{Physics Department, University of Minnesota, 116 Church Street S.E., Minneapolis, MN 55455}
\altaffiltext{\STScI}{Space Telescope Science Institute, 3700 San Martin
Dr., Baltimore, MD 21218}
\altaffiltext{\SAIC}{Liberal Arts Department, 
School of the Art Institute of Chicago, 
112 S Michigan Ave, Chicago, IL 60603}
\altaffiltext{\Yale}{Department of Physics, Yale University, P.O. Box 208210, New Haven,
CT 06520-8120}
\altaffiltext{\LLNL}{Institute of Geophysics and Planetary Physics, Lawrence
Livermore National Laboratory, Livermore, CA 94551}
\altaffiltext{\BCCP}{Berkeley Center for Cosmological Physics,
Department of Physics, University of California, and Lawrence Berkeley
National Labs, Berkeley, CA 94720}

\def\Michigan{1}
\def\Munich{2}
\def\ExcellenceCluster{3}
\def\CfA{4}
\def\KICPChicago{5}
\def\PhysicsUChicago{6}
\def\UChicago{7}
\def\UPenn{8}
\def\Harvard{9}
\def\EFIChicago{10}
\def\IAPFrance{11}
\def\Miss{12}
\def\AAUChicago{13}
\def\ANL{14}
\def\NIST{15}
\def\PUC{16}
\def\McGill{17}
\def\Berkeley{18}
\def\UFlorida{19}
\def\Colorado{20}
\def\NASA{21}
\def\Davis{22}
\def\LBNL{23}
\def\Caltech{24}
\def\Arizona{25}
\def\MIT{26}
\def\MPE{27}
\def\CaseWestern{28}
\def\Minnesota{29}
\def\STScI{30}
\def\SAIC{31}
\def\Yale{32}
\def\LLNL{33}
\def\BCCP{34}

\author{J.~Song\altaffilmark{\Michigan},
A.~Zenteno\altaffilmark{\Munich,\ExcellenceCluster},
B.~Stalder\altaffilmark{\CfA},
S.~Desai\altaffilmark{\Munich,\ExcellenceCluster},
L.~E.~Bleem\altaffilmark{\KICPChicago,\PhysicsUChicago},
K.~A.~Aird\altaffilmark{\UChicago},
R.~Armstrong\altaffilmark{\UPenn},
M.~L.~N.~Ashby\altaffilmark{\CfA},
M.~Bayliss\altaffilmark{\CfA,\Harvard}, 
G.~Bazin\altaffilmark{\Munich,\ExcellenceCluster},
B.~A.~Benson\altaffilmark{\KICPChicago,\EFIChicago},
E.~Bertin\altaffilmark{\IAPFrance},
M.~Brodwin\altaffilmark{\Miss},
J.~E.~Carlstrom\altaffilmark{\KICPChicago,\PhysicsUChicago,\EFIChicago,\AAUChicago,\ANL}, 
C.~L.~Chang\altaffilmark{\KICPChicago,\EFIChicago,\ANL}, 
H.~M. Cho\altaffilmark{\NIST}, 
A.~Clocchiatti\altaffilmark{\PUC},
T.~M.~Crawford\altaffilmark{\KICPChicago,\AAUChicago},
A.~T.~Crites\altaffilmark{\KICPChicago,\AAUChicago},
T.~de~Haan\altaffilmark{\McGill},
M.~A.~Dobbs\altaffilmark{\McGill},
J.~P.~Dudley\altaffilmark{\McGill},
R.~J.~Foley\altaffilmark{\CfA}, 
E.~M.~George\altaffilmark{\Berkeley},
D.~Gettings\altaffilmark{\UFlorida},
M.~D.~Gladders\altaffilmark{\KICPChicago,\AAUChicago},
A.~H.~Gonzalez\altaffilmark{\UFlorida},
N.~W.~Halverson\altaffilmark{\Colorado},
N.~L.~Harrington\altaffilmark{\Berkeley},
F.~W.~High\altaffilmark{\KICPChicago,\AAUChicago}, 
G.~P.~Holder\altaffilmark{\McGill},
W.~L.~Holzapfel\altaffilmark{\Berkeley},
S.~Hoover\altaffilmark{\KICPChicago,\EFIChicago},
J.~D.~Hrubes\altaffilmark{\UChicago},
M.~Joy\altaffilmark{\NASA},
R.~Keisler\altaffilmark{\KICPChicago,\PhysicsUChicago},
L.~Knox\altaffilmark{\Davis},
A.~T.~Lee\altaffilmark{\Berkeley,\LBNL},
E.~M.~Leitch\altaffilmark{\KICPChicago,\AAUChicago},
J.~Liu\altaffilmark{\Munich,\ExcellenceCluster},
M.~Lueker\altaffilmark{\Berkeley,\Caltech},
D.~Luong-Van\altaffilmark{\UChicago},
D.~P.~Marrone\altaffilmark{\Arizona},
M.~McDonald\altaffilmark{\MIT},
J.~J.~McMahon\altaffilmark{\Michigan},
J.~Mehl\altaffilmark{\KICPChicago,\AAUChicago},
S.~S.~Meyer\altaffilmark{\KICPChicago,\PhysicsUChicago,\EFIChicago,\AAUChicago},
L.~Mocanu\altaffilmark{\KICPChicago,\AAUChicago},
J.~J.~Mohr\altaffilmark{\Munich,\ExcellenceCluster,\MPE},
T.~E.~Montroy\altaffilmark{\CaseWestern},
T.~Natoli\altaffilmark{\KICPChicago,\PhysicsUChicago},
D.~Nurgaliev\altaffilmark{\Harvard}, 
S.~Padin\altaffilmark{\KICPChicago,\AAUChicago,\Caltech},
T.~Plagge\altaffilmark{\KICPChicago,\AAUChicago},
C.~Pryke\altaffilmark{\Minnesota}, 
C.~L.~Reichardt\altaffilmark{\Berkeley},
A.~Rest\altaffilmark{\STScI},
J.~Ruel\altaffilmark{\Harvard},
J.~E.~Ruhl\altaffilmark{\CaseWestern}, 
B.~R.~Saliwanchik\altaffilmark{\CaseWestern}, 
A.~Saro\altaffilmark{\Munich},
J.~T.~Sayre\altaffilmark{\CaseWestern}, 
K.~K.~Schaffer\altaffilmark{\KICPChicago,\EFIChicago,\SAIC}, 
L.~Shaw\altaffilmark{\McGill,\Yale},
E.~Shirokoff\altaffilmark{\Berkeley,\Caltech}, 
R.~\v{S}uhada\altaffilmark{\Munich},
H.~G.~Spieler\altaffilmark{\LBNL},
S.~A.~Stanford\altaffilmark{\Davis,\LLNL},
Z.~Staniszewski\altaffilmark{\CaseWestern},
A.~A.~Stark\altaffilmark{\CfA}, 
K.~Story\altaffilmark{\KICPChicago,\PhysicsUChicago},
C.~W.~Stubbs\altaffilmark{\CfA,\Harvard}, 
A.~van~Engelen\altaffilmark{\McGill},
K.~Vanderlinde\altaffilmark{\McGill},
J.~D.~Vieira\altaffilmark{\KICPChicago,\PhysicsUChicago,\Caltech},
R.~Williamson\altaffilmark{\KICPChicago,\AAUChicago}, 
and
O.~Zahn\altaffilmark{\Berkeley,\BCCP}
}

\begin{abstract}

We present the results of the ground- and space-based optical and
near-infrared (NIR) follow-up of 224 galaxy cluster candidates
detected with the Sunyaev-Zel'dovich (SZ) effect in the
720 deg$^2$ of the South Pole Telescope (SPT) survey completed in the 2008 and
2009 observing seasons.  We use the optical/NIR data to establish whether each candidate is associated with an overdensity of galaxies and to estimate the cluster redshift. 
Most photometric redshifts are derived through a combination of three different cluster redshift estimators using red-sequence galaxies, resulting in an accuracy of $\Delta z/(1+z)=0.017$, determined through comparison with a subsample of 57 clusters for which we have spectroscopic redshifts.  We successfully measure redshifts for 158 systems and present redshift lower limits for the remaining candidates.  The redshift distribution of the confirmed clusters extends to $z=1.35$ with a median of $\zmed=0.57$.  Approximately 18\% of the sample with measured redshifts lies at $z>0.8$.  We
  estimate a lower limit to the purity of this SPT SZ-selected sample
  by assuming that all unconfirmed clusters are noise fluctuations in
  the SPT data.  We show that the cumulative purity at detection significance $\xi>5$ ( $\xi>4.5$) is $\ge 95\%$ ($\ge 70\%$).  
  We present the red brightest cluster galaxy (rBCG) positions for the sample and examine the offsets between the SPT candidate position and the rBCG.  The radial distribution of offsets is similar to that seen in X-ray-selected cluster samples, providing no evidence that SZ-selected cluster samples include a different fraction of recent mergers than X-ray-selected cluster samples.

\end{abstract}



\bigskip

\section{Introduction}
\label{sec:intro}

In November 2011, the South Pole Telescope \citep[SPT;][]{carlstrom11} 
collaboration completed a 2500 \degs\ 
survey, primarily aimed at detecting distant, massive galaxy clusters
through their 
Sunyaev-Zel'dovich (SZ) effect signature.  In \citet[][R12 hereafter]{reichardt12c}, the SPT team presented a catalog of 224 
cluster candidates from 720 \degs\ observed in the 2008-2009 seasons.  In this work, we present 
the optical and near-infrared (NIR) follow-up observations of the cluster candidates
reported in R12, mainly focusing on 
follow-up strategy, 
confirmation and empirical purity estimate for the cluster candidates, photometric redshift estimations of confirmed clusters, and the spatial position of the red brightest cluster galaxies.

Galaxy clusters have long been used for the study of structure
formation and cosmology \citep[e.g.,][]{geller82,white93b}.
Soon after the discovery of the cosmic acceleration
\citep{schmidt98,perlmutter99a}, it became clear that measurements of
the redshift evolution of the cluster mass function could provide a
powerful tool to further understand the underlying causes
\citep{wang98,haiman01,holder01b,battye03}.  More precise theoretical investigations \citep[][]{majumdar03,hu03b,majumdar04,molnar04,wang04,lima05,lima07} identified the key challenges associated with cluster surveys, which include: (1) producing large uncontaminated samples selected by an observable property that is closely related to the cluster mass, (2) measuring cluster redshifts for large samples and (3) precisely calibrating the cluster masses.

Competitive approaches to producing large cluster samples include optical multiband 
surveys \citep[e.g.,][]{gladders05,koester07a}, infrared surveys \citep[e.g.,][]{eisenhardt08,muzzin08,papovich08}, X-ray surveys 
\citep[e.g.,][]{finoguenov07,pacaud07,vikhlinin09,finoguenov10,mantz10a,fassbender11,lloyddavies11,suhada12},
 and millimeter-wave (mm-wave) surveys	 
 \citep{vanderlinde10,marriage11b,planck11-5.1a,reichardt12c}.  The
 mm-wave surveys capitalize on the cluster SZ
effect signature, which is produced by the inverse Compton scattering
of cosmic microwave background photons by the energetic electrons
within the cluster \citep{sunyaev72}.  The surface brightness of the SZ effect 
is redshift-independent, making SZ surveys a particularly powerful tool for 
identifying the most distant clusters.  
It is typical for X-ray and mm-wave surveys to have accompanying multiband optical imaging to enable photometric redshift measurements; these multiband optical data also enable a second stage of cluster candidate
confirmation, verifying the purity estimation of the X-ray or SZ-selected
cluster samples.

Ideally, one would coordinate an SZ survey with a deep, multiband
optical survey over the same region; indeed, the Dark Energy Survey
\citep[DES\footnote{www.darkenergysurvey.org};][]{cease08,mohr08} and  the SPT are coordinated in this way.  
Because of the different development timelines for the two projects,
it has been necessary to undertake extensive cluster-by-cluster
imaging follow-up for SPT using a series of ground-based telescopes together with
space-based NIR imaging (from \spitzer\ and \wise).  
The NIR data are
of particular importance in the confirmation and redshift estimation
of the $z>1$ massive galaxy clusters, which are especially interesting for both
cosmological studies and studies of the evolution of clusters
themselves.
Pointed observations were used in \citet[][H10 hereafter]{high10} to provide redshift and richness estimates of the SZ detections of \citet{vanderlinde10}, and subsequently by \citet{williamson11} and \citet{story11}.  

Cluster samples from high-resolution SZ surveys can be also used to explore the evolution of cluster properties as a function of redshift.  
Previous studies using X-ray-selected clusters have identified a correlation between the dynamical state of a cluster and the projected offset between the X-ray centers and the brightest cluster galaxy \citep[e.g.,][]{katayama03,sanderson09,haarsma10,fassbender11,stott12}.  In principle, SZ-selected clusters can serve as laboratories to search for this correlation also, if the spatial resolution of SZ detections is high enough to detect the significance of offsets between the SZ centers and the brightest cluster galaxies.  Systematic comparison between X-ray and SZ samples will indicate if the selection of the two methods differs in terms of the dynamical state of clusters.

This paper is structured as follows: we briefly describe the SPT data
and methods for extracting the cluster sample in \S\ref{sec:spt
summary}.  In \S\ref{sec:followup}, we provide details of the
follow-up strategy, as well as data processing.  \S\ref{sec:method} is
dedicated to a detailed description of the analysis of our follow-up
data, including redshift estimation using optical and \spitzer \ data,
the derivation of redshift lower limits for those systems that are
not confirmed, and the selection of red brightest cluster galaxies (rBCGs) in the clusters.  
Results are presented in \S\ref{sec:results} and discussed further in \S\ref{sec:conclusion}.  Throughout this paper,
we use the AB magnitude system for optical and NIR observations unless otherwise
noted in the text.

\section{Discovery \& Followup}

\subsection{SPT Data}
\label{sec:spt summary}

Here we briefly summarize
the analysis of the SPT data and the extraction of cluster candidates
from that data; we refer the reader to R12 and previous SPT cluster
publications \citep{staniszewski09,vanderlinde10,williamson11} for
more details.

The SPT operates in three frequency bands, although only data from the
95~GHz and 150~GHz detectors were used in finding clusters.  The data
from all detectors at a given observing frequency during an observing
period (usually 1-2 hours) are combined into a single map.  The data
undergo quality cuts and are high-pass filtered and
inverse-noise-weighted before being combined into a map.  Many
individual-observation maps of every field are co-added (again using
inverse noise weighting) into a full-depth map of that field, and the
individual-observation maps are differenced to estimate the map noise.
The 95~GHz and 150~GHz full-depth maps of a given field are then combined
using a spatial-spectral matched filter \citep[e.g.,][]{melin06} that
optimizes signal-to-noise on cluster-shaped objects with an SZ
spectral signature.  Cluster candidates are identified in the
resulting filtered map using a simple peak detection algorithm, and
each candidate is assigned a signal-to-noise value based on the peak
amplitude divided by the RMS of the filtered map in the neighborhood
of the peak.  Twelve different matched filters are used, each assuming
a different scale radius for the cluster, and the maximum
signal-to-noise for a given candidate across all filter scales is
referred to as $\xi$, which we use as our primary SZ observable.  In
2008, the 95~GHz detectors in the SPT receiver had significantly lower
sensitivity than the 150~GHz array, and the cluster candidates from
those observations are identified using 150~GHz data only; the
candidates from 2009 observations were identified using data from both
bands.  The data from the two observing seasons yielded a total of 224
cluster candidates with $\xi$$\ge$$4.5$---the sample discussed here.

\subsection{Optical/NIR Imaging}
\label{sec:followup}

The cluster candidates detected using the method described above
are followed up by optical and, in many cases, NIR instruments.  In
this section, we describe the overall optical/NIR follow-up strategy,
the different imaging and spectroscopic observations and facilities
used, and the data reduction methods used to process the raw images to
catalogs.

\subsubsection{Imaging Observations}

The optical/NIR follow-up strategy has evolved since the first SPT-SZ
candidates were identified.  Originally we imaged regions
of the sky with uniformly deep, multiband observations in $griz$ optical bands to
confirm SZ detections and estimate redshifts as in \citet{staniszewski09}.  For the first SPT cluster 
candidates, we used imaging from the Blanco Cosmology Survey \citep[BCS;][]{desai12}
to follow up candidates in parts of the 2008 fields.  The BCS is a
60-night, $\sim$$80$~$\degs$ NOAO survey program carried out in
2005-2008 using the Blanco/MOSAIC2 $griz$ filters.  The BCS survey was
completed to the required depths for $5\sigma$ detection at $0.4L^*$
within $2\farcs3$ apertures up to $z$$\sim$$1$.  The goal of this
survey was to provide optical imaging over a limited area of the SPT
survey to enable rapid optical follow-up of the initial SPT survey fields.

For clusters outside the BCS region we initially obtained deep $griz$ imaging on a 
cluster-by-cluster basis.  But as the SPT survey proceeded and the cluster candidate list grew, 
it became clear that this strategy was too costly, given the limited access to follow-up time.  
Moreover, eventually the full SPT region will be imaged to uniform
10$\sigma$ depths of mag$\sim$24 in $griz$ by the DES.  We therefore
switched to an adaptive strategy of follow-up in which we observed
each SPT cluster candidate to the depth required to find an optical counterpart 
and determine its redshift.

For each SPT cluster candidate, we perform an initial pre-screening of candidates using the
Digitized Sky Survey (DSS)\footnote{http://archive.stsci.edu/dss/}.  We examine DSS images using 3 bands\footnote{http://gsss.stsci.edu/SkySurveys/Surveys.htm} for each cluster
candidate to determine whether it is ``low-$z$'' or ``high-$z$,''
where the redshift boundary lies roughly at $z=0.5$.  We find that this visual classification identifies spectroscopically and photometrically confirmed SPT clusters out to $z$=0.5 in the DSS photographic plates.  We use the DSS
designation to prioritize the target list for the appropriate telescope, instrument and
filters with which we observe each candidate.  Specifically,
candidates that are clearly identified in DSS images are 
likely to be low-$z$ clusters and are designated
for follow-up observations on small-aperture (1m-2m) telescopes.  Otherwise, candidates are classified as 
high-$z$ candidates and therefore designated for large-aperture (4m-6.5m)
telescopes.   The various ground- and space-based facilities used to collect
optical/NIR imaging data on SPT clusters are summarized in
Table~\ref{tab:oir cameras}.  Each telescope/instrument combination is
assigned a numeric alias that is used to identify the source of
the redshift data for each cluster in Table~\ref{tab:master list}.

For the $\ge 4$m-class observations, we use an adaptive
filter and exposure time strategy so that we can efficiently bracket the cluster member galaxy's 4000\AA\ break to the depth required for redshift estimation.  In this approach we start with a first imaging pass,
where each candidate is observed in the $g$, $r$, and $z$ bands to
achieve a depth corresponding to a $5\sigma$ detection of a $0.4L^*$~galaxy at $z=0.8$, $\sim$23.5 mag and 21.8 mag in $r$ and $z$ bands respectively, based on the \citet{bruzual03} red-sequence model (for more details about the model, see \S~\ref{sec:method:photoz}).  Observations are also taken in a single blue filter for photometric calibration using the stellar locus (discussed in \S\ref{sec:reduction}).  For candidates with no obvious optical counterpart after first-pass observations, a second-pass is executed to get to $z=$~0.9, $\sim$23 mag and 22 mag in $i$ and $z$ bands respectively.  

If the candidate is still not confirmed after the second-pass in $i$ and $z$ bands, and is not covered by the \spitzer/IRAC pointed observations described below, we attempt to obtain ground-based NIR imaging for that candidate using the NEWFIRM camera on the CTIO Blanco telescope.  The data presented here are imaged with NEWFIRM during three observing runs in 2010 and 2011, yielding $K_s$ data for a total of 31 candidates.  Typical observations in $K_s$ consist of 16 point dither patterns, with 60 second exposures divided among 6 coadds at each dither position.  Median seeing during the 2010 runs was 1$\farcs$05; during the 2011 run observing conditions were highly variable and the seeing ranged from $1\farcs05$ to 2$\farcs6$ with median seeing $\sim1\farcs2$.

We note that most of the galaxy cluster candidates in this work with significance $\xi>4.8$ were imaged with \spitzer\ \citep{werner04}.   More specifically, \spitzer/IRAC \citep{fazio04} imaging has been obtained for 99 SZ cluster candidates in this work.  The on-target \spitzer \ observations consist of 8$\times100$\,s and 6$\times30$\,s dithered exposures at 3.6$\mu$m and 4.5$\mu$m, respectively. The deep 3.6$\mu$m observations should produce $5\sigma$ detections of passively-evolving $0.1L^*$ cluster galaxies at $z = 1.5$ ($\sim$17.8 mag (Vega) at $z=1.5$).

For some of the NIR analysis, we augment the data from our
\spitzer \ and NEWFIRM observations with the recently released all-sky
Wide-field Infrared Survey Explorer \citep[\wise;][]{wright10} data.  Finally, we note that a few of the clusters were observed with Magellan/Megacam to obtain weak gravitational lensing mass measurements
\citep{high12}. These data are naturally much deeper than our initial followup imaging.

\begin{deluxetable*}{llllllcl}
\tabletypesize{\scriptsize}
\tablecaption{Optical and infrared imagers\label{tab:oir cameras}}
\tablewidth{0pt}
\tablehead{
\colhead{Ref.\tablenotemark{a}} &
\colhead{Site} &
\colhead{Telescope} &
\colhead{Aperture} &
\colhead{Camera} &
\colhead{Filters\tablenotemark{b}} &
\colhead{Field} &
\colhead{Pixel scale} \\
\colhead{~} &
\colhead{~} &
\colhead{~} &
\colhead{(m)} &
\colhead{~} &
\colhead{~} &
\colhead{~} &
\colhead{($\arcsec$)}
}
\startdata
1 & Cerro Tololo & Blanco & 4 & MOSAIC-II & $griz$ & $36\arcmin \times 36\arcmin$ & $0.27$ \\
2 & Las Campanas & Magellan/Baade & 6.5 & IMACS f/2 & $griz$ & $27\farcm4 \times 27\farcm4$ & $0.200$ \\
3 & Las Campanas & Magellan/Clay & 6.5 & LDSS3 & $griz$ & $8\farcm3$ diam.\ circle  & $0.189$ \\
4\tablenotemark{c} & Las Campanas & Magellan/Clay & 6.5 & Megacam & $gri$ & $25\arcmin \times 25\arcmin$ & $0.16$ \\
5 & Las Campanas & Swope & 1 & SITe3 & $BVRI$ & $14\farcm8 \times 22\farcm8$ & $0.435$ \\
6 & Cerro Tololo & Blanco & 4 & NEWFIRM & $K_s$ & $28\arcmin \times 28\arcmin$ & $0.4$ \\
7 & \nodata & Spitzer Space Telescope & 0.85 & IRAC & 3.6$\mu$m, 4.5$\mu$m & $5\farcm2 \times 5\farcm2$ & $1.2$ \\
8 & \nodata & Wide-field Infrared Survey Explorer & 0.40 &  \nodata &  3.4$\mu$m, 4.6$\mu$m & $47\arcmin \times 47\arcmin$ & $6$ \\

\enddata
\tablecomments{Optical and infrared cameras used in SPT follow-up observations. }
\tablenotetext{a}{Shorthand alias used in Table \ref{tab:master list}.}
\tablenotetext{b}{Not all filters were used on every cluster.}
\tablenotetext{c}{Megacam data were acquired for a large follow-up weak-lensing program.}
\end{deluxetable*}

\subsubsection{Data Processing}
\label{sec:reduction}

We use two independent optical data processing systems.  One system, which we refer to as the PHOTPIPE pipeline,
is used to process all optical data except Magellan/Megacam data, and the other, which is a development version of the Dark Energy Survey data management (DESDM) system, is used only to process the Blanco/Mosaic2 data.  PHOTPIPE  was used to process optical data for previous SPT cluster catalogs \citep{vanderlinde10,high10,williamson11}; the DESDM system has been used as a cross-check in these works and was the primary reduction pipeline used in \citet{staniszewski09}.  

The basic stages of the PHOTPIPE pipeline, initially developed for the SuperMACHO and ESSENCE
projects and described in \citet{rest05a}, \citet{garg07}, and \citet{miknaitis07}, include flat-fielding, astrometry, coadding, and source extraction.  Further details are given in H10.   In the DESDM system \citep{ngeow06,mohr08,desai12}, the data from each night first undergo detrending corrections, which include cross-talk correction, overscan correction, trimming, bias subtraction, flat fielding and illumination correction.  Single epoch images are not remapped to avoid correlating noise, and so we also perform a pixel-scale correction that brings all sources on an image to a common photometric zeropoint.  For $i$ and $z$ bands we also carry out a fringe correction.  Astrometric calibration is done by using the AstrOmatic code {\tt SCAMP}~\citep{bertin06} and the USNO-B catalog.  Color terms to transform to the SDSS system rely on photometric solutions derived from observations of SDSS equatorial fields during photometric nights \citep{desai12}.

In both pipelines, coaddition is done using {\tt SWarp}~\citep{bertin02}.  In the DESDM system the single epoch images contributing to the coadd are brought to a common zeropoint using stellar sources common to pairs of images.  The final photometric calibration of the coadd images is carried out using the stellar color-color locus as a constraint on the zeropoint offsets between neighboring bands \citep[e.g.,][]{high09}, where the absolute photometric calibration comes from 2MASS \citep{skrutskie06}.  For $griz$ photometry the calibration is done with reference to the median SDSS stellar locus \citep{covey07}, but for the Swope data using Johnson filters, the calibration relies on a stellar locus derived from a sequence of models of stellar atmospheres from PHOENIX \citep{brott05} with empirically measured CCD, filter, and atmosphere responses.  Cataloging is done using SExtractor \citep{bertin96}, and within the DESDM catalogs we calibrate {\it mag\_auto} using stellar locus.    

Quality checks of the photometry are carried out on a cluster by cluster basis using the scatter of stars about the expected stellar locus and the distribution of offsets in the single-epoch photometry as a function of calibrated magnitude (so-called photometric repeatability tests).  Poor quality data or failed calibrations are easily identified as those coadds with high stellar locus scatter and or high scatter in the photometric repeatability tests \citep[see][]{desai12}.

NEWFIRM imaging data are reduced using the FATBOY pipeline \citep{eikenberry06}, originally developed for the FLAMINGOS-2 instrument, and modified to work with NEWFIRM data in support of the Infrared Bootes Imaging Survey \citep{gonzalez10}.  Individual processed frames are combined using {\tt SCAMP} and {\tt SWarp}, and photometry is calibrated to 2MASS.

\spitzer/IRAC data are reduced following the method of \citet{ashby09}.  Briefly, we correct for column pulldown, mosaic the individual exposures, resample to 0\farcs86 pixels (half the solid angle of the native IRAC pixels), and reject cosmic rays.  Magnitudes are measured in 4\arcsec--diameter apertures and corrected for the 38\% and 40\% loss at 3.6$\mu$m and 4.5$\mu$m respectively due to the broad PSF \citep[see Table 3 in ][]{ashby09}.  The {\it Spitzer} photometry is crucial to the measurement of photometric redshifts for clusters at $z \ga 0.8$, as described in \S\ref{sec:method:photoz}.

The acquisition and processing for the
initial weak lensing Megacam data is described in detail in \citet{high12}. 
These data are reduced separately from the other imaging
data using the Smithsonian Astrophysical
Observatory (SAO) Megacam reduction pipeline.  Standard raw CCD image processing, cosmic-ray
removal, and flat-fielding are performed, as well as an additional
illumination correction to account for a low-order scattered light
pattern.  The final images are coadded onto a single pixel grid with a
pixel scale of $0\farcs16$ using {\tt SWarp}.  Sources are detected in
the coadded data in dual-image mode using {\tt SExtractor}, where the
$r$ band image serves as the detection image.  The photometry is
calibrated by fitting colors to the stellar locus, and color-term
corrections are accounted for in this step.  The color term is roughly $0.10(g-i)$ for the $g'$ band, $-0.02(g-i)$ for the $r'$ band, and $-0.03(g-i)$ for the $i'$ band.

\subsection{Spectroscopic Observations}
\label{sec:spec}

We have targeted many of the SPT clusters with long-slit and
multi-object spectroscopy, and some of the spectroscopic redshifts
have appeared in previous SPT publications.  We have used a variety of
instruments: GMOS-S\footnote{http://www.gemini.edu/node/10625} on
Gemini South, FORS2 \citep{appenzeller98} on VLT, LDSS3 on
Magellan-Clay, and 
the IMACS camera on Magellan Baade
(in long-slit mode and with the
GISMO\footnote{http://www.lco.cl/telescopes-information/\\magellan/instruments/imacs/gismo/gismoquickmanual.pdf} complement).

A detailed description of the configurations, observing runs and
reductions will be presented elsewhere (J.\ Ruel et al., in prep).
For a given cluster we 
target bright galaxies that lie on the clusters' red sequence and observe these galaxies with a
combination of filter and disperser that yields a low-resolution
spectrum around their  \ion{Ca}{2} H\&K lines and 4000\AA \ break. CCD reductions are
made using standard packages, including COSMOS \citep{kelson03} for
IMACS data and IRAF\footnote{http://iraf.noao.edu} for GMOS and FORS2.  Redshift measurements
are made by cross-correlation with the RVSAO package \citep{kurtz98}
and a proprietary template fitting method that uses SDSS DR2
templates.  Results are then visually confirmed using strong spectral
features.

In Table~\ref{tab:specz}, the source for every spectroscopic
redshift is listed, along with the number of cluster members used in deriving the redshift.  
For clusters for which we report our own spectroscopic measurements, 
we list an instrument name and observation date; we give 
a literature
reference for those for which we report a value from the
literature.
In Table \ref{tab:master list}, we report spectroscopic redshifts for
57 clusters, of which 36 had no previous spectroscopic redshift in the
literature. Unless otherwise noted, the reported cluster redshift is
the robust biweight average of the redshifts of all spectroscopically
confirmed cluster members, and the cluster redshift uncertainty is
found from bootstrap resampling.  

\begin{center}
\begin{deluxetable*}{lcccc |  lcccc}
\tabletypesize{\scriptsize}
\tablecaption{Spectroscopic Follow-Up\label{tab:specz}}
\tablewidth{0pt}
\tablehead{\colhead{Cluster} &\colhead{Inst} & \colhead{Obs} & \colhead{\#} & \colhead{Refs} &\colhead{Cluster}&\colhead{Inst} & \colhead{Obs} & \colhead{\#} & \colhead{Refs}}
\startdata
SPT-CLJ0000-5748        &  GMOS-S & Sep 2010 & 26 &   --                &        SPT-CLJ2058-5608        &  GMOS-S & Sep 2011  &  9 &         -- \\
SPT-CLJ0205-5829        &  IMACS    & Sep 2011 & 9   &          a          &        SPT-CLJ2100-4548  &  FORS2 & Aug 2011 & 19 &  -- \\     
SPT-CLJ0205-6432        &  GMOS-S & Sep 2011 & 15 &                      &        SPT-CLJ2104-5224  &  FORS2 & Jun 2011 & 23 &   -- \\    
SPT-CLJ0233-5819        &  GMOS-S & Sep 2011 & 10 &                      &       SPT-CLJ2106-5844  &  FORS2 & Dec 2010 & 15 &  -- \\     
SPT-CLJ0234-5831        &  GISMO   & Oct 2010  & 22 &          b           &                                               &  GISMO & Jun 2010 & 3 &    i\\
SPT-CLJ0240-5946        &  GISMO   & Oct 2010  & 25 &         --           &       SPT-CLJ2118-5055  &  FORS2 & May 2011 & 25 &  --\\      
SPT-CLJ0254-5857        &  GISMO   & Oct 2010  & 35 &          b           &      SPT-CLJ2124-6124  &  GISMO & Sep 2009 & 24 &   --\\     
SPT-CLJ0257-5732        &  GISMO   & Oct 2010  & 22 &         --           &       SPT-CLJ2130-6458  &  GISMO & Sep 2009 & 47 &   --\\     
SPT-CLJ0317-5935        &  GISMO   & Oct 2010  & 17 &         --           &      SPT-CLJ2135-5726 &  GISMO & Sep 2010 & 33 &  --\\       
SPT-CLJ0328-5541        &     --           &       --   & -- &    c                        &         SPT-CLJ2136-4704  &  GMOS-S & Sep 2011 & 24 & -- \\     
SPT-CLJ0431-6126        &     --           &       --   & --  &   c                        &        SPT-CLJ2136-6307  &  GISMO & Aug 2010 & 10 & --\\       
SPT-CLJ0433-5630        &  GISMO     &  Jan 2011  &  22 &     --         &         SPT-CLJ2138-6007  &  GISMO & Sep 2010 & 34 &   -- \\    
SPT-CLJ0509-5342        &  GMOS-S  &  Dec 2009 &  18 &      d, e     &        SPT-CLJ2145-5644  &  GISMO & Sep 2009 & 37 &   --\\     
SPT-CLJ0511-5154        &  GMOS-S  &  Sep 2011 &  15 &        --       &     SPT-CLJ2146-4633  &  IMACS &  Sep 2011 &  17 &  --\\    
SPT-CLJ0516-5430        &  GISMO     &  Sep 2010 &   48 &      f          &        SPT-CLJ2146-4846  &  GMOS-S  & Sep 2011  & 26 & --\\    
SPT-CLJ0521-5104        &     --           &  --        & -- &      e                      &        SPT-CLJ2148-6116  &  GISMO  & Sep 2009  & 30 & --\\     
SPT-CLJ0528-5300        & GMOS-S   &  Jan 2010 &  20 & d, e           &     SPT-CLJ2155-6048  &  GMOS-S  & Sep 2011  & 25 & --\\    
SPT-CLJ0533-5005        &  LDSS3     &  Dec 2008  & 4   &       d        &       SPT-CLJ2201-5956  &         --      &       --      &       -- & c\\    
SPT-CLJ0534-5937        &  LDSS3     &  Dec 2008  & 3   &     --          &      SPT-CLJ2300-5331  &  GISMO  & Oct 2010  & 24 &   --\\   
SPT-CLJ0546-5345        &  GISMO    &  Feb 2010  & 21 &       g         &      SPT-CLJ2301-5546  &  GISMO &  Aug 2010  & 11 & --\\     
                                            &  GMOS-S  &  Dec 2009 & 2   & e                &      SPT-CLJ2331-5051  &  GMOS-S & Aug 2010 & 28 & --\\      
SPT-CLJ0551-5709        &  GISMO    &  Sep 2010  & 34 &        d        &                                               &  GISMO  &  & 50 &                d \\
SPT-CLJ0559-5249        &  GMOS-S  & Nov 2009  &  37 &      d, e     &     SPT-CLJ2332-5358  &  GISMO  &  Jul 2009 &   24 &   --  \\       
SPT-CLJ2011-5725        &     --           &  --       & --     &   f                       &     SPT-CLJ2337-5942  &  GMOS-S &  Aug 2010  & 19 & d \\    
SPT-CLJ2012-5649        &     --           &  --       &--  & c                            &       SPT-CLJ2341-5119  &  GMOS-S &  Aug 2010  & 15 & d \\    
SPT-CLJ2022-6323        &  GISMO    &  Oct 2010  &  37 &      --          &    SPT-CLJ2342-5411  &  GMOS-S  &  Sep 2010  & 11 & -- \\  
SPT-CLJ2023-5535        &     --          &   -- & --  &          f                         &      SPT-CLJ2351-5452   &        --       &      --       &  -- &    h \\    
SPT-CLJ2032-5627        &  GISMO    & Oct 2010   &  31 &      --          &       SPT-CLJ2355-5056  &  GISMO &   Sep 2010 &   37 &    --\\        
SPT-CLJ2040-5725        &  GISMO    & Aug 2010  &  5   &      --          &      SPT-CLJ2359-5009   &  GISMO &  Aug 2010 & 21 &     --\\ 
SPT-CLJ2043-5035        &  FORS2    & Aug 2011  &  21 &       --        &                                               &  GMOS-S  &  Dec 2009 &  9 &  -- \\
SPT-CLJ2056-5459        &  GISMO    & Aug 2010  &  12 &       --        &                       &                              &             &               &        \\
\enddata
\tablecomments{Instruments [Inst] :  GMOS-S on Gemini South 8m, IMACS on Magellan Baade 6.5m, GISMO complement to IMACS on Magellan Baade 6.5m, LDSS3 on Magellan Clay 6.5, FORS2 on VLT Antu 8m;  Observing dates [Obs]: dates each data taken; Number of galaxies [\#]: Number of galaxies used in deriving redshifts; References [Refs]:  $^a$\citet{stalder12}, $^b$\citet{williamson11}, $^c$\citet{struble99},  $^d$\citet{high10}, $^e$\citet{sifon12}, $^f$\citet{boehringer04}, $^g$\citet{brodwin10}, $^h$\citet{buckleygeer11}, $^i$\citet{foley11} }
\end{deluxetable*}
\end{center}

\section{Methodology}
\label{sec:method}

In this section, we describe the analysis methods used to: 1) extract  
cluster redshift estimates and place redshift limits; 2) empirically verify the
estimates of catalog purity; and 3) measure rBCG positions.

\subsection{Photometric Redshifts}
\label{sec:method:photoz}

Using the procedure described in \S\ref{sec:followup}, we obtain
ground-based imaging data and galaxy catalogs that in most cases allow
us to identify an obvious overdensity  of red-sequence galaxies 
within approximately an arcminute of the SPT candidate position.  For
these optically confirmed cluster candidates, we proceed to estimate a photometric redshift.

In this work, we employ three methods (which we refer to Method 1, 2, and 3 in the following sections) to estimate cluster redshifts
from optical imaging data.  Two methods (Method 1 and 2) use the color of the galaxies in
the cluster red sequence, and the third (Method 3) uses the average
of red-sequence galaxy photometric redshifts estimated with a
neural-network algorithm, trained with the magnitudes of similar
galaxies.  In the optical analysis for our two previous
cluster catalog releases \citep{high10,williamson11},  we relied on Method 1 for the results and Method 3 as a cross-check.  In this work, we improve the
precision of the measured redshifts by applying
multiple redshift estimation algorithms and combining the results.
Through cross-checks during the analysis, we find that these methods 
have different failure modes and that comparing the results
provides a way of identifying systems that require additional
attention (including systems where the cluster's central region is contaminated by foreground stars or the cluster resides in a crowded field).

All three methods use the single-stellar-population (SSP) models of
\citet[BC03;][]{bruzual03}.  These models allow us to 
transform the location of the red-sequence overdensity in color
space to a redshift estimate.  A model for the red galaxy population
as a function of redshift is
built assuming a single burst of star formation at $z_f=3$ followed by
passive evolution thereafter.  Models are selected over a range of
metallicities and then calibrated to reproduce the color and tilt of
the red sequence in the Coma cluster \citep{eisenhardt07} at $z=0.023$.
The calibration procedure is described in more detail in
\citet{song12a}.  The red sequence model prescribed in a similar way has been demonstrated to adequately describe the bright end of the cluster red sequence \citep[][]{blakeslee03,tran07,muzzin09,high10,mancone10,stott10,song12a}.  These models are used in determining exposure
times and appropriate filter combinations for imaging observations,
and in the calculation of redshifts and redshift limits from those
observations.

\subsubsection{Photometric Redshift Measurement Methods}
\label{subsec:photoz}

In Method 1, a cluster is confirmed by identifying an excess of
galaxies with colors consistent with those derived from BC03 (simultaneously for all observed filters), after subtracting the background surface density.  The background-subtracted galaxy number is extracted from an aperture
within a radius of (3.5,2.5,1.5)$'$ from the SPT candidate position
and uses galaxies with photometric color uncertainties
$\le$(0.25,0.35,0.45) and apparent magnitudes brighter than
$m^*$+(3,2,1) (or the magnitude limit of the data) in the red
sequence based on the same BC03 models, for $z$$<$$0.2$, $0.2$$<$$z$$<0.6$, and $z$$>$$0.6$
respectively.  The background measurement is obtained by applying the
same criteria outside of the cluster search aperture.  The redshift is
estimated from the most significant peak in this red-sequence galaxy
excess.  Improvements over the implementation in H10 include using additional
colors ($r$-$z$ and $g$-$i$, plus NIR colors) in the red-sequence
fitting, using the deeper photometry available from coadded images,
and sampling the entire CCD mosaic rather than a single CCD for better
background estimation.

Method 2 is similar in that it searches for an overdensity
of red-sequence galaxies.  This method, used to estimate the redshifts for a
sample of 46 X-ray-selected clusters \citep{suhada12}, is described
and tested in more detail in \citet{song12a}.  It includes a measure
of the background surface density based on the entire imaged sky area
surrounding each cluster candidate and subtracts the background from the red galaxy counts in an aperture of 0.8~Mpc.  
Only galaxies with luminosity $>0.4$L$^*$ and magnitude uncertainty $\le0.25$ 
are used, and
the aperture and luminosity are recalculated for each potential redshift.
Originally as described in \citet{song12a}, we search for an overdensity of red-sequence galaxies using two or three available color-redshift
combinations simultaneously for every cluster; essentially, we scan
outwards in redshift using the following color combinations: $g$-$r$
and $g$-$i$ for $z<$$0.35$, $g$-$i$ and $r$-$i$ and $r$-$z$ for
$0.3$$<$$z$$<$$0.75$, $r$-$z$ and $i$-$z$ for $z$$>$$0.75$.  The cluster
photometric redshift is extracted from the peak of the galaxy
overdensity in redshift space.  The redshift is 
 then refined by fitting the red-sequence
overdensity distribution in redshift space with a Gaussian function.
The version used here (which is the same as the method used in
\citealt{suhada12}) has one more refinement, in which the colors of the galaxies that
lie in the peak redshift bin identified by the overdensity method are
converted into individual galaxy photometric redshifts.
In this conversion we assume that the galaxies are red-sequence
cluster member galaxies, and the photometric redshift uncertainty
reflects the individual photometric color errors.  A final cluster
redshift is calculated as an inverse-variance-weighted mean of these galaxy
photometric redshifts.

Method 3 shares the same principle as the other two in that it
involves searching for a density peak in the galaxy distribution near
the position of the SPT candidate.  We first select individual red cluster
members using location relative to the SPT candidate position and
galaxy color as the criteria for cluster membership.  For the redshifts presented here, this is done visually
using pseudo-color coadded images for each cluster, although in
principle this could be automated.  Galaxy selection is \emph{not} confined by
a specific radial distance from SZ centers as in the other two
methods, nor by photometric uncertainties.   Selected galaxies are then fed
into Artificial Neural Networks \citep[ANN$z$;][]{collister04}, which is trained using the same BC03 models used in the other methods.   ANN$z$ returns redshift estimates for individual galaxies, 
and a peak in galaxy redshift  distribution is adopted as the initial cluster
redshift.   Then, as in Method 2, individual galaxy photometric redshifts are averaged using inverse-variance
weighting to produce the cluster photometric redshift.   With this initial estimate of the redshift, we then perform an outlier rejection using iterative $1\sigma$ clipping, where the $1\sigma$ corresponds to the root-mean-square (RMS) variation of the measured galaxy photometric redshift distribution.  Once the rejection is carried out, we refine the cluster photometric redshift estimate using the weighted mean of the non-rejected sample of cluster galaxies.  No outlier rejection is
undertaken if there are fewer than 20 selected galaxies in the
original sample.  

Method 3 is a good cross-check, as well as a stand-alone redshift estimator, because we can visually  confirm which galaxies contribute to the redshift determination.  Although this method requires photometry in more than just two bands, it appears to be less susceptible to the problems in two-band methods that are associated with pileup of red sequence galaxies at redshifts where the 4000\AA\  break is transitioning out of a band.  

\begin{figure}[htb]
  \begin{center}
  \includegraphics[scale=0.5]{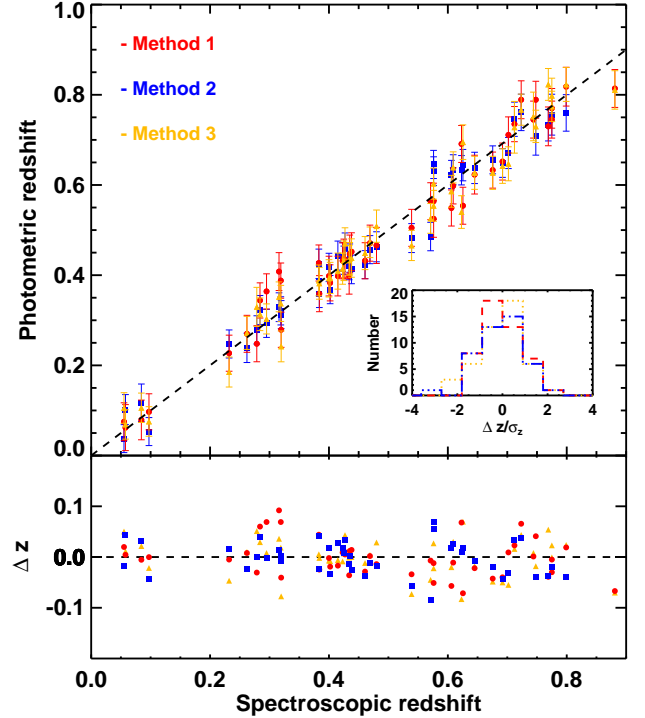}
  \caption{$\it{Top}$: Photometric redshift \zphot\ versus spectroscopic redshift \zspec\  for each redshift estimation method for 47 spectroscopically confirmed clusters at $z<0.9$ where we use only $griz$ photometry.  $\it{Bottom}$: the distribution of the photometric redshift residuals $\Delta z = \zphot - \zspec$ as a function of \zspec.    $\it{Inset}$: the normalized residual distributions, which all have $RMS(\Delta z/\sigma_{\zphot})\sim1$.   The RMS scatter of $\Delta z/(1+z)$ is 0.028, 0.023 and 0.024 for Methods 1 (red dot and red dashed line), 2 (blue square and blue dash-dot-dot line) and 3 (yellow triangle and yellow dotted line), respectively.}
  \label{fig:photoz individual performance}
  \end{center}
\end{figure}

Next we characterize redshift estimates from each method using spectroscopically confirmed clusters.  We use 47 clusters with spectroscopic redshifts (\zspec) where only $griz$ data are used for photometric redshift estimation.  In this process, photometric redshift (\zphot) biases (namely, smooth trends of photometric redshift offset as a function of redshift) are measured and corrected in Method 1 and 2, while no significant bias correction is necessary for Method 3.   Bias corrections depend on several factors, such as filters used for data, redshift of clusters, and the depth of the data.  They are separately measured in those different cases per method as a function of (1+$z$) at a level of 0.01-0.03 in redshift for clusters with redshift measured in $griz$ filters, \spitzer-only, and $BVRI$ filters at $z>0.5$.  The largest bias correction is needed for clusters observed from SWOPE using $BVRI$ filters with maximum correction of 0.13 at around $z\sim0.4$ where the filter transitions from $B$-$V$ to $V$-$R$ occurs to capture the red sequence population.  This affects two clusters in the final sample.  Once biases are removed, we examine the photometric-to-spectroscopic redshift offsets to characterize the performance of each method.  We find the RMS in the quantity $\Delta z$/(1+\zspec), where $\Delta z = \zphot - \zspec$ to have values of 0.028, 0.023, and 0.024 in Methods 1, 2, and 3, respectively (see Figure~\ref{fig:photoz individual performance}).  We note that some of the bias and systematic error, especially at higher redshift, could be due to the mismatch between the SEDs in the red sequence model and the cluster population, which could arise from variations in star formation history or AGN activity.

Our goal is not only to estimate accurate and precise cluster
redshifts, but also to accurately characterize the uncertainty in these
estimates.  To this end, we use the spectroscopic
subsample of clusters to estimate a systematic floor \sigsys \ in addition to 
the statistical component.  We do this by requiring
that the reduced $\chi^2$ describing the normalized
photometric redshift deviations from the true redshifts $\chired=\sum
(\Delta z/\sigma_{\zphot})^2 / \ndof$ have a value $\chired\sim1$ for
each method, where $\sigma_{\zphot}$ is the uncertainty in measured
$\zphot$ and $\ndof$ is the number of degrees of freedom.  We adopt
uncertainties $\sigma_{\zphot}^2=\sigstat^2+\sigsys^2$ and adjust $\sigsys$ to obtain the correct $\chired$.   
In this tuning process we also include redshift estimates of the same cluster from multiple
instruments when that cluster has been observed multiple times.  This
allows us to test the performance of our uncertainties over a broader
range of observing modes and depths than is possible if we just use
the best available data for each cluster.

For Method 1 we separately measure the systematic floor \sigsys \ for
each different photometric band set. For the $grizK_S$ instruments
(Megacam, IMACS, LDSS3, MOSAIC2, NEWFIRM), we estimate $\sigsys=0.039$
; for the $BVRI$ instrument (Swope), $\sigsys=0.033$; and for
\spitzer-only, $\sigsys=0.070$.  In Method 2, we find $\sigsys=0.030$
for the $griz$ instruments (Megacam, IMACS, LDSS3, MOSAIC2).  For
Method 3 we estimate $\sigsys=0.028$ for the $griz$ instruments.

Once this individual estimation and calibration is done, we conduct an
additional test on the redshift estimation methods, again using the
spectroscopic subsample. The purpose of this test is to see how the
quality of photometry (i.e., follow-up depth) affects the estimations.
We divide the spectroscopic sample into two groups: in one group, the
photometric data is kept at full depth, while the photometric data in
the other group is manually degraded to resemble the data from the
shallowest observations in the total follow-up sample.  To create the
`shallow' catalogs, we add white noise to the full-depth coadds and then extract and calibrate catalogs from these artificially noisier images.  Results of this test show that the accuracy of the
photometric redshift estimation is affected by the poorer photometry,
but that this trend is well captured by the statistical
uncertainties in each estimation method.

\subsubsection{Combining Photo-z Estimates to Obtain \zcomb}
\label{sec:method:combine}

Once redshifts and redshift uncertainties are estimated with each method independently, we compare the different redshift estimates of the same cluster.  Note that this comparison is not possible for Swope or \spitzer-only redshifts, which are measured only with Method 1.  Outliers at $\ge 3 \sigma$ ($>$6\%) in 1+\zphot \ are identified for additional inspection.  In some cases, there
is an easily identifiable and correctable issue with one of the methods, such as misidentification of cluster members.  If, however, it is not possible to identify an obvious problem, the outliers are excluded from the combining procedure.  This outlier rejection, which occurs only in two cases, causes less than 0.05 change in the combined \zphot in both cases.  

We combine the individual estimates into a final best
redshift estimate, $z_\mathrm{comb}$, using inverse-variance weighting
and accounting for the covariance between the methods, 
which we expect to be non-zero given the similarities in the methods and the common data used.  Correlation coefficients for the photometric redshift errors among the different methods are measured
using the spectroscopic sample. The measured correlation coefficient,
$r_{ij}$, between each pair of methods is 0.11 (Method 1 \& 2), 0.40
(Method 2 \& 3) and 0.19 (Method 1 \& 3).

With the correlation coefficients we construct the optimal combination
of the individual estimates as:
\begin{equation}
z_\mathrm{comb} = \frac{1}{\sum_{ij} W_{ij}} \sum_{i} \sum_{j} W_{ij} z_j,
\end{equation}
where $W_{ij} = C^{-1}_{ij}$, and the covariance matrix $C_{ij}$ is
comprised of the square of the individual uncertainties along the
diagonal elements ($\sigma_i^2$) and the product of the measured
correlation coefficient and the two individual uncertainty components
(i.e., $r_{ij}$$\sigma_i$$\sigma_j$) on the off-diagonal elements.
The associated uncertainty is
\begin{equation}
\sigma^2_{z_\mathrm{comb}} = \frac{1}{\sum_{ij} W_{ij}}.
\end{equation}

\begin{figure}[htb]
  \begin{center}
  \includegraphics[scale=0.5]{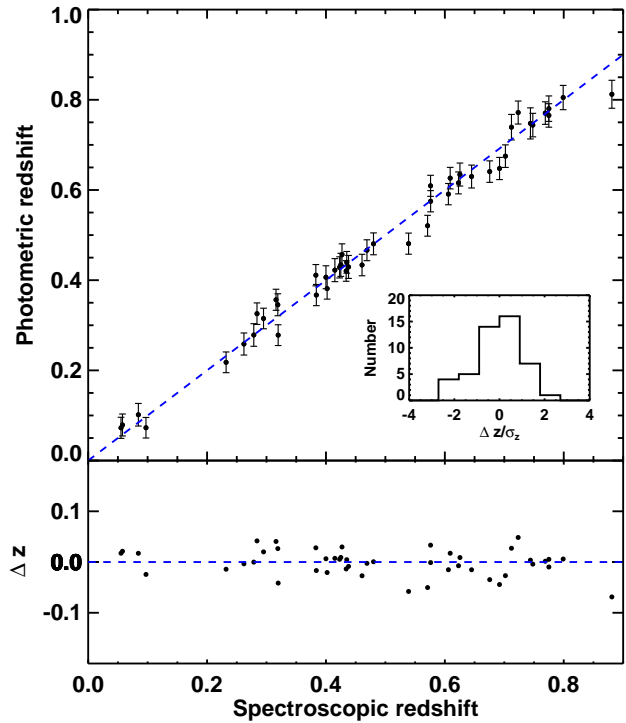}
  \caption{$\it{Top}$: Weighted mean photometric redshift \zcomb \ versus ~spectroscopic redshift using the same sub-sample as in Figure~\ref{fig:photoz individual performance}.   
$\it{Bottom}$: the distribution of the redshift errors.  The RMS scatter in $\Delta z/(1+\zspec)$=0.017.  $\it{Inset}$: histogram of the normalized redshift error distribution,
 which is roughly Gaussian with RMS $\simeq 1$.}
  \label{fig:bestz performance}
  \end{center}
\end{figure}

Because of the positive correlations between the three
methods' errors, the errors on \zcomb \ are larger than would be the
case for combining three independent estimates; however, we do see an
improvement in the performance of the combined redshifts relative to
the individual estimates that is consistent with the expectation given
the correlations.  The performance of this combined redshift method is
presented in Figure~\ref{fig:bestz performance}; the residual
distribution is roughly Gaussian, and the associated uncertainties
provide a good description of the scatter of the redshift estimates
about the spectroscopic redshifts (the RMS variation of $\Delta
z/\sigma_{\zphot}$ is 1.04).  The benefit from combining different
measurements is evidenced from the tighter distribution in the
redshift versus $z_\mathrm{spec}$ plot; the RMS scatter of $\Delta
z/(1+\zspec)$ is 0.017, corresponding to a $\sim$40\% improvement in the accuracy
relative to the accuracy of a single method.  

\subsubsection{\spitzer\ Photometric Redshifts}
\label{sec:spitzerz}

For clusters where we do not have deep enough optical data to estimate
a redshift but that do have \spitzer\ coverage, we use the algorithm used in Method 1 to
measure the redshifts using \spitzer-only colors in the same manner as
we do with optical data.  Overdensities of red galaxies in
clusters have been identified using \spitzer-only color selection at high
redshift, where the IRAC bands are probing the peak of the stellar
emission \citep{stern05,papovich08}, rather than bracketing the 4000\AA \ break.  Note that the concerns about the impact of recent star formation or AGN activity on photometric redshift estimates are not as serious in the IRAC bands as in the optical bands, because the portion of the spectrum probed is less sensitive to these potential sources of contamination.
In our sample, the comparison of \spitzer-only redshifts with
spectroscopically derived redshifts shows good performance, indicating
that the assumption of a well-developed red-sequence appears to hold
out to $z \ga 1$ \citep[e.g.,][]{bower92,eisenhardt08,muzzin09}.  Note that
the possibility of the cluster being at lower redshift is already
ruled out from the available optical data for these candidates.

\begin{figure}[htb]
  \begin{center}
  \includegraphics[scale=0.5]{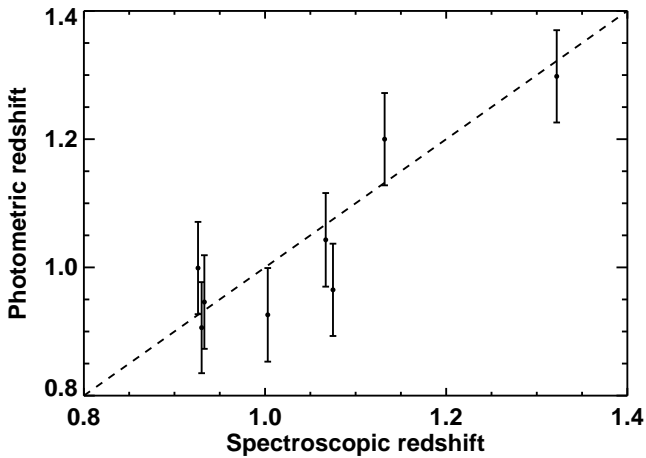}
  \caption{Photometric redshift vs. \zspec \ for clusters
    where only the \spitzer\ IRAC 3.6$\mu$m-4.5$\mu$m colors are used.
    In all cases where we present \spitzer\ photo-z's, we have optical
    data to rule out the presence of a low-redshift cluster.}
  \label{fig:spitzer only photoz}
  \end{center}
\end{figure}

Figure~\ref{fig:spitzer only photoz} shows the performance of the
\spitzer-only redshifts in eight clusters where spectroscopic
redshifts are available.  Although the accuracy in \zphot \ is lower ($\Delta
z/(1+z)\approx 0.049$) than those derived from optical-only or
optical-IRAC colors, the performance is reliable.  We flag
these cases in the final table to make note of this difference in
method.  The larger uncertainties of \spitzer--only derived redshifts
are possibly due to the broad width of IRAC filters and the fact that AGN emission or vigorous star formation can shift the location of the 1.6$\mu$m bump. 

\subsubsection{Redshift Limits}
\label{sec:method:zlim}
In most cases there is an obvious, rich overdensity of red
cluster galaxies in our follow-up imaging, from which it is
straightforward to confirm the galaxy cluster giving rise to the SZ
signal and to estimate the cluster redshift.  
For unconfirmed candidates, 
it is not possible to say with absolute certainty that no
optical/NIR counterpart exists; with real, finite-depth optical and
NIR data, the possibility always exists that the cluster is distant enough that
no counterpart would have been detected at the achieved optical/NIR
depth.  Assigning a relative probability to these two interpretations
of an optical/NIR non-detection (i.e., a false detection in the SZ
data or a higher-redshift cluster than the optical/NIR observations
could detect) is especially important for interpreting the SZ cluster
sample cosmologically.  To this end, we calculate a lower redshift
limit for every SZ-selected candidate for which no counterpart has
been found.  Because the optical/NIR follow-up data is not homogeneous,
we do this separately for each unconfirmed candidate.

To estimate the depth of the optical/NIR coadded images, we utilize a
Monte-Carlo based technique described in \citet{ashby09}. In brief, we
perform photometry of the sky in various apertures at 1500 random
positions in each image. To measure the sky noise, we then fit a
Gaussian function to the resulting flux distribution (excluding the
bright tail which is biased by real sources in the image).  Taking the
measured sky noise from 3\arcsec-diameter apertures, we add a
PSF-dependent aperture correction.  A redshift limit is derived for
each filter by matching a 0.4$L^*$ red-sequence galaxy from the BC03 model to the measured 10$\sigma$
magnitude limit.  We use the redshift limit from the second deepest
filter with regards to 0.4$L^*$ red-sequence objects,
as we require a minimum of two filters to measure a
redshift.  These redshift limits are compared to 
limits derived by comparing observed number counts of galaxies as a function of magnitude
to distributions derived from
much deeper data \citep{zenteno11}.
We find the two independent redshift limit estimations are in good agreement.

For cluster candidates with \spitzer/IRAC observations, the redshift
estimation is not limited by the depth of the optical data, and we use
the IRAC data to calculate a lower redshift limit for these
candidates.  The IRAC data are highly uniform, with depth sufficient
to extract robust photometry down to $0.1 L^*$ out to a redshift of
$z=1.5$.  In principle, $\sim 0.5 L^*$ photometry should be sufficient
for redshift estimation; however, we adopt $z=1.5$ as a conservative
lower redshift limit for any unconfirmed candidates with IRAC data.

\subsection{NIR Overdensity Estimates for Unconfirmed Candidates}
\label{sec:method:nirrich}

For cluster candidates for which we are unable to estimate a redshift,
we can in principle go beyond a simple binary statement of
``confirmed/unconfirmed" using NIR data.  Even if there is not a
sufficient number of galaxies in the NIR data to estimate a red
sequence, there is information in the simple overdensity of objects
(identified in a single NIR band) within a certain radius of an SPT
candidate, and we can use this information to estimate the probability
of that candidate being a real, massive cluster.  We can then use this
estimate to sharpen our estimate of the purity of the SPT-selected
cluster sample.  We calculate the single-band NIR overdensity for all
unconfirmed candidates using \wise\ data, and we compare that value to
the same statistic estimated on blank-field data.  We perform the same
procedure using \spitzer/IRAC and NEWFIRM data for unconfirmed
candidates that were targeted with those instruments.  For comparison,
we also calculate the same set of statistics for each confirmed
cluster above $z=0.7$.

We estimate the galaxy overdensity within a $1\ \mathrm{arcmin}^2$
aperture.  To increase the signal-to-noise of the estimator, we assume
an angular profile shape for the cluster galaxy distribution and fit
the observed distribution to this shape.
The assumed galaxy density profile is a projected $\beta$ model with
$\beta=1$ (the same profile assumed for the SZ signal in the
matched-filter cluster detection algorithm in R12).  We have tried
using a projected NFW profile as well, and the results do not change
in any significant way (due to the relatively low signal-to-noise in
the NIR data).  The central amplitude,
background amplitude, scale radius, and center position (with respect to the center of the SZ signal in SPT data) are free parameters in the fit.  The number of galaxies above background within $1\
\mathrm{arcmin}^2$---which we will call \sigoneam---is then calculated
from the best-fit profile.  The same procedure is repeated on
fields not expected to contain massive galaxy clusters,
and the value of \sigoneam \ for every SPT candidate is compared to
the distribution of \sigoneam \ values in the blank fields.  The key
statistic is the fraction of blank fields that had a \sigoneam \ value
larger than a given SPT candidate, and that value is recorded as \pblank \ in
Table~\ref{tab:master list} for every high-redshift ($z \ge 0.7$) or
unconfirmed candidate.  This technique, including using the
blank-field statistic as the primary result, is similar to the
analysis of WISE data in the direction of unconfirmed \planck \ Early
SZ clusters in \citet{sayers12}, although that analysis used raw
galaxy counts within an aperture rather than profile fitting.

The model fitting is performed using a simplex-based $\chi^2$
minimization, with any parameter priors enforced by adding a $\chi^2$
penalty.  The positional offset $\chi^2$ penalty is $\Delta \chi^2 =
(\Delta \theta / \sigma_{\Delta \theta})^2$, where $\sigma_{\Delta
  \theta}$ is chosen to be 0.25\arcmin, based on the SZ/BCG offset
distribution in Figure~\ref{fig:offset_distribution} and the value of
$r_{200}$ for a typical-mass SPT cluster at high redshift.\footnote{$r_{200}$
is defined as the radius within which the average density is 200 times the 
mean matter density in the universe.}
A prior is
enforced on the scale radius from below and above by adding $\chi^2$
penalties of $(\theta_s/\theta_{s,\mathrm{max}})^2$ and
$(\theta_{s,\mathrm{min}}/\theta_s)^2$, where
$\theta_{s,\mathrm{max}}$ is chosen to be 0.75\arcmin based on the
$\theta_s$ distribution in known high-redshift SPT clusters with NIR
data, and $\theta_{s,\mathrm{min}}$ is chosen to be 0.125\arcmin to
prevent the fitter from latching onto small-scale noise peaks.

For \spitzer/IRAC and \wise, the fit is performed on the 3.6$\mu$m
and the 3.4$\mu$m data, respectively; for NEWFIRM, the fit is performed on the $K_s$-band data.  For both
\spitzer/IRAC and NEWFIRM, a single magnitude threshold is used for every candidate; this threshold is
determined by maximizing the signal-to-noise on the \sigoneam \
estimator on known clusters while staying safely away from the
magnitude limit of the shallowest observations.  The \spitzer/IRAC
data is very uniform, and the 3.6$\mu$m magnitude threshold chosen is
18.5 (Vega).  The NEWFIRM $K_s$-band data is less uniform, but a
magnitude threshold of $18$ (Vega) is safe for all observations.  For
these instruments (IRAC and NEWFIRM), the blank fields on which the
fit is performed come from the \spitzer \ Deep Wide-Field Survey
(SDWFS) region \citep{ashby09}, which corresponds to the Bootes field
of the NOAO Deep Wide-Field Survey \citep[NDWFS;][]{jannuzi99}.  The depth of the
SDWFS/NDWFS observations for both instruments (19.8 in Vega for
\spitzer/IRAC 3.6$\mu$m and
19.5 in Vega for NEWFIRM $K_s$) is more than sufficient for our chosen
magnitude thresholds.

For \wise, in which the non-uniform sky coverage results in
significant variation in magnitude limits, we perform the blank-field
fit on data in the immediate area of the cluster (within a $\sim 20^\prime$
radius).  Under the assumption that the \wise\ magnitude limit does
not vary over this small an angular scale, we use all detected
galaxies brighter than 18th magnitude (Vega) in both the cluster and
blank-field fits.

\subsection{Identifying rBCGs in SPT Clusters}
\label{sec:method:bcg id}

An rBCG in this work is defined as the brightest galaxy among the red-sequence galaxies for each cluster.  We employ the terminology rBCGs, rather than BCGs, to allow for the rare possibility of an even brighter galaxy with significant amounts of ongoing star formation, because the selection is restricted by galaxy colors along the cluster red sequence.  We visually inspect pseudo-color images built with the appropriate filter combinations (given the cluster redshift) around the SZ candidate position.  We search a region corresponding to
the projected cluster virial region, defined by $\theta_{200}$, given
the mass estimate from the observed cluster SPT significance $\xi$ and
photometric redshift \zphot.  

There are 12 clusters out of the 158 with measured photometric redshifts (excluding the candidates with redshift limits) that are excluded from the rBCG selections.  Eight of those are excluded due to contamination by a bright star that obscures more than one third of the area of the 3$\sigma$ SPT positional uncertainty region.  Another cluster is excluded due to a bleed trail making the rBCG selection ambiguous, and three other clusters are excluded due to a high density in the galaxy population that, given the delivered image quality, makes it impossible to select the rBCG.
\section{Results}
\label{sec:results}
The complete list of 224 SPT cluster candidates with SZ detection
significance $\xi \ge 4.5$ appears in Table~\ref{tab:master list}.  
The table includes SZ cluster candidate positions on the sky [RA Dec], SZ detection significance [$\xi$], and spectroscopic redshift [\zspec] when available.  For confirmed clusters, the table includes photometric redshift and uncertainty [\zcomb$\pm$$\sigma_{z_\mathrm{comb}}$], estimated as described in \S\ref{sec:method:photoz}.  Unconfirmed candidates are assigned redshift lower limits, estimated as described in \S\ref{sec:method:zlim}.  

We also report a redshift quality flag for each \zphot\ in Table~\ref{tab:master list}.  For most of the confirmed clusters with reliable photometric redshift measurements, we set Flag = 1.  There is one cluster (SPT-CL~J2146-4846) for which the three individual photometric redshifts are not statistically consistent ($\ga 3\sigma$ outliers) for which we set Flag = 2.  We still report the combined redshift for that cluster as in other secure systems.  We have 6 cases where
we only use Swope + Method 1, and 25 cases where we only use \spitzer \ +
Method 1 for redshift estimation, both cases marked with Flag = 3.  We note that the photometric redshift bias correction for two clusters (SPT-CL~J0333-5842 and SPT-CL~J0456-6141) is at a higher level than the typical bias correction on other clusters (see \ref{subsec:photoz} for more detail on the bias correction.  There are 2 cases (SPT-CL~J0556-5403 and SPT-CL~J0430-6251) where
we quote only a Method 1 redshift even for MOSAIC or IMACS data, marked with Flag = 4.  
In the coadded optical images for SPT-CL~J0556-5403, we identify an overdensity of faint red galaxies at the location of the SPT candidate. This optical data is too shallow, however, to allow for secure redshift estimation, but we are able to measure a redshift by combining this data with NEWFIRM imaging.  This cluster is the only candidate where we rely on photometric redshift from $i$-$K_s$.  SPT-CL~J0430-6251
is in a field very crowded with large scale structure, making redshift estimation difficult.

In the Appendix, we
discuss certain individually notable candidates---such as associations
with known clusters that appear to be random superpositions and
candidates with no optical/NIR confirmation but strong evidence from
the NIR overdensity statistic.

\begin{figure}[htb]
  \begin{center}
  \includegraphics[scale=0.44]{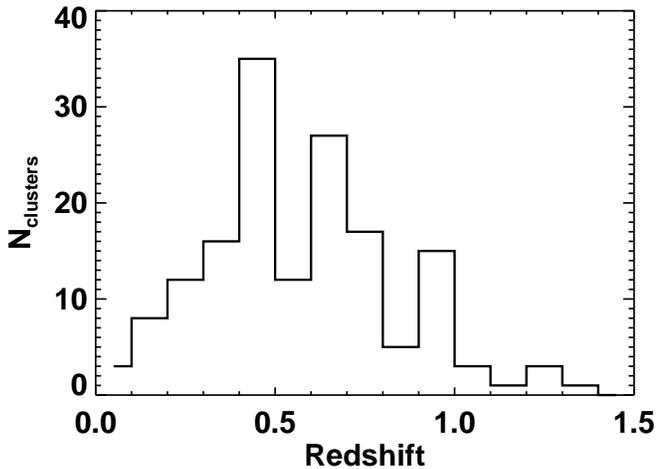}
  \caption{Redshift histogram of 158 confirmed clusters, in bins of $\Delta z = 0.1$.  Note that
    about 18\% of the total sample comes from $z \ge 0.8$.}  
      \label{fig:zdistribution}
  \end{center}
\end{figure}

\subsection{Redshift Distribution}
\label{sec:result:zdist}

The redshift distribution for the 158 confirmed
clusters is shown in Figure~\ref{fig:zdistribution}.  The median redshift is $z=0.57$, with 28 systems ($\sim$18\% of
the sample) lying at $z > 0.8$.  The cluster
with the highest photometric redshift is SPT-CL~J2040-4451 at $z=1.35\pm0.07$
(estimated using \spitzer/IRAC data) and the highest-redshift spectroscopically confirmed cluster is SPT-CL~J0205-5829 at $z=1.32$.  (This cluster is discussed in detail in \citealt{stalder12}.)     

The high fraction of SPT clusters at $z>0.8$ is a consequence of the 
redshift independence of the SZ surface brightness and the arcminute 
angular resolution of the SPT, which is well-matched to the angular size of 
high-$z$ clusters.  X-ray surveys, in contrast, are highly efficient at finding
nearby clusters, but the mass limit of an X-ray survey will increase with 
redshift due to cosmological dimming.
ROSAT cluster surveys lack the sensitivity
to push to these high redshifts except in the
deepest archival exposures.  XMM--Newton archival
surveys \citep[e.g.,][]{lloyddavies11,fassbender11} and coordinated surveys of
contiguous regions \citep[e.g.,][]{pacaud07,suhada12} have sufficient sensitivity
to detect systems like those found by SPT, but the solid
angle surveyed is currently smaller.  For example, the
\citet{fassbender11} survey for high redshift clusters will eventually
cover approximately 80~deg$^2$, whereas the mean sky density of the
SPT high-redshift and high-mass systems is around one every
25~deg$^2$.  Therefore, one would have expected the
\citet{fassbender11} XDCP survey to have found around three clusters of
comparable mass to the SPT clusters, which is in fact consistent with
their findings.  The vast majority of the high redshift X-ray-selected
sample available today is of significantly lower mass than SPT
selected samples, simply because the X-ray surveys do not yet cover
adequate solid angle to find these rare, high mass systems.

Clusters samples built from NIR galaxy catalogs have an even higher fraction of high-redshift
systems than SZ-selected samples---for example, 
in the IRAC Shallow Cluster Survey \citep[ISCS;][]{eisenhardt08} a sample of 335 clusters has been identified out to $z \sim 2$, a third of which are at $z>1$.  However, the typical ISCS cluster mass is $\sim10^{14}$ M$_\odot$ \citep{brodwin07}, significantly lower than the minimum mass of the SPT high-redshift sample.  As with the X-ray selected samples, the \spitzer\ sample includes some massive clusters, including the recently discovered IDCS J1426.5+3508 at $z=1.75$, which was subsequently also detected in the SZ \citep{stanford12,brodwin12}.  However, the \spitzer\ surveys to date do not cover the required solid angle to find these massive systems in the numbers being discovered by SPT.

\begin{figure}[htb]
  \begin{center}
  \includegraphics[scale=0.48]{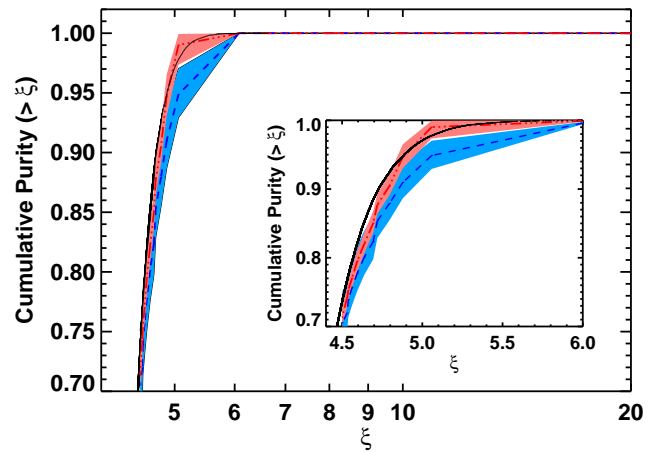}
  \caption{Cumulative purity estimates derived from the optical/NIR followup  compared to simulated purity predictions (solid black line).  The inset plot is zoomed-in to the $\xi$ range between 4.5 and 6.0 and binned more finely.  The purity is calculated from the follow-up confirmation rate: 1) (dashed blue) assuming all clusters without a clear optical or NIR counterpart are false SZ detections (i.e., 100\% optical completeness) and all optical confirmations are robust (100\% optical purity); and 2) (dash-dot-dot red) assuming, as justified in the text, 97\% optical completeness and 96\% optical purity but taking into
account clusters confirmed through other means such as X-ray observations.  1$\sigma$ uncertainties in the purity estimates from follow-up are shown with shaded blue or red regions (see Section~\ref{sec:result:contamination}). 
}
  \label{fig:purity}
  \end{center}
\end{figure}

\subsection{Purity of the SPT Cluster Candidates}
\label{sec:result:contamination}

For a cluster sample to be useful for cosmological purposes, it is important to know the purity of the sample, defined as
\begin{equation}
f_\mathrm{pure} = \frac{N_\mathrm{real}}{N_\mathrm{tot}} =  1- \frac{N_\mathrm{false}}{N_\mathrm{tot}},
\end{equation}
where $N_\mathrm{tot}$ is the total number of cluster candidates,
$N_\mathrm{real}$ is the number of candidates corresponding to real
clusters, and $N_\mathrm{false}$ is the number of false detections.
For an SZ-selected cluster sample with reasonably deep and complete
optical/NIR follow-up, a first-order estimate of $N_\mathrm{real}$ is
simply the number of candidates with successfully estimated redshifts.
In Figure~\ref{fig:purity}, we show two estimations of purity for the
720~deg$^2$ SPT-SZ sample; the first in blue, assuming that all
cluster candidates with no redshift measurements are noise
fluctuations, and the other in red, taking into account incompleteness of our follow-up data.  
The blue/red shaded regions in the figure
correspond to the 1$\sigma$ uncertainties on the purity, estimated from Poisson
noise on $N_\mathrm{false}$ for the blue region and as described below for the
red region.
We also show the expected purity, estimated from the
total number of candidates in the sample presented in this work combined with the
false detection rate from the simulations used to test the SZ cluster
finder (R12 Figure 1).  

The possibility of real clusters beyond the redshift reach of our
optical/NIR redshift estimation techniques makes the blue line in
Figure~\ref{fig:purity} a lower limit to the true purity of the
sample.  As discussed in \S \ref{sec:method:nirrich}, we use
single-band NIR data to estimate the probability that each unconfirmed
candidate is a ``blank field'', i.e., a field with typical or
lower-than-typical NIR galaxy density.  Candidates with no optical/NIR
confirmation but with a low blank field probability \pblank, are
potential high-redshift systems that merit further follow-up study.  These systems
can also give an indication of how much we underestimate our sample
purity when we assume any optical/NIR non-confirmation is a spurious
SPT detection.  By definition, a low \pblank \ implies some NIR
overdensity towards the SPT detection, but perhaps not large enough to be an SPT-detectable cluster.  We can roughly calibrate the \pblank \
values to SPT detectability by investigating the results of the NIR
overdensity estimator on solidly confirmed, high-redshift SPT
clusters.  There are 19 clusters with spectroscopic redshifts above
$z=0.7$, and the average \spitzer/IRAC \pblank \ value for these
clusters is $0.04$, while the average WISE \pblank \ value is $0.05$.
Only three of these clusters have NEWFIRM data, and the average
NEWFIRM \pblank \ value is $0.07$.  Only one cluster in this high-$z$
spectroscopic sample has an IRAC $\pblank > 0.1$, while three have
WISE $\pblank > 0.1$.  So a rough threshold for SPT-type clusters
appears to be $\pblank \le 0.1$.  We have nine unconfirmed cluster
candidates that meet this criterion in at least one of the NIR
catalogs, including five that are at $\pblank \le 0.05$.  If we
assumed all of the $\pblank \le 0.05$ clusters were real, it would
imply that the completeness of the optical/NIR redshift estimation was
$\sim 97\%$, i.e., we have 163 real clusters of which we were able to
estimate redshifts for 158.

Conversely, the possibility of false associations of spurious SZ
detections with optical/NIR overdensities would act in the other
direction.  Tests of one of the red-sequence methods on blank-field
data produced a significant red-sequence detection on approximately $4\%$ of fields without SPT detected clusters.  Assuming that the cross-checks with other methods
would remove some of these, we can take this as an upper limit to this
effect.

We therefore provide a second estimate of purity from the optical/NIR confirmation rate,
taking into
account the possibility of real clusters for which we were unable to
successfully estimate a redshift (redshift completeness $<100\%$) and
spurious optical/NIR associations with SPT noise peaks (redshift
purity $<100\%$).  
From the above arguments, we assume 97\% for
redshift completeness and 96\% for redshift purity.  
For each value of SPT significance $\xi$, 
we use binomial
statistics to ask how often a sample of a given purity with total candidates $N(>\xi)$
would produce the observed number of successful optical/NIR redshift estimates 
$N_\mathrm{conf}(>\xi)$,
given the redshift completeness and false rate.  
An extra constraint
is added to this calculation based on data independent of the
optical/NIR imaging that confirms many of the SPT candidates as real,
massive galaxy clusters.  Specifically, we assume an SPT candidate is
a real, massive cluster independent of the optical/NIR imaging data
(and remove the possibility of that candidate being a false
optical/NIR confirmation of an SPT noise peak) if: 1) it is associated 
with a ROSAT Bright Source Catalog source; 2) we have obtained X-ray
data in which we can confirm a strong, extended source; or 3) we have
obtained spectroscopic data and measured a velocity dispersion for
that system.  The red solid line and shaded region in
Figure~\ref{fig:purity} show the maximum-likelihood value and $68\%$
limits for the true purity of the SPT sample under these assumptions.

The purity measured in this work is in good agreement with the model
for the SPT sample purity.  In particular, all clusters with $\xi>6$
have identified optical counterparts with photometric redshift
estimates.  This is consistent with the expectation of the model and a
demonstration that the SPT selected galaxy cluster sample is
effectively uncontaminated at $\xi>6$.
With decreasing significance, the number of
noise fluctuations in the SPT maps 
increases compared to the number of real clusters on the sky, and the purity
decreases.  
The cumulative purity of the sample is $\sim$70\% above $\xi=4.5$ 
and reaches $\sim$100\% above $\xi=5.9$.  Of course, if one requires
optical confirmation in addition to the SPT detection, then the sample is effectively 100\% pure over the full sample at $\xi>4.5$.  

=We note that there is no significant difference in 
false detection rate (based on optical confirmation alone) between 
cluster candidates selected with 150~GHz data alone and those detected with the 
multiband strategy (see \S\ref{sec:spt summary} for details).  Roughly 
1/4 of the survey area was searched for clusters using 150~GHz data only, 
and in that area we have 12 unconfirmed candidates, including one 
above $\xi=5$; in the 3/4 of the area selected using multiband data, we have
54 unconfirmed candidates, including five above $\xi=5$.  These totals are 
consistent within $1 \sigma$ Poisson uncertainties.

The high purity of the SPT selected cluster sample is comparable to
the purity obtained in previous X-ray cluster surveys
\citep[i.e.][]{vikhlinin98,mantz08,vikhlinin09}, indicating that these
intracluster medium based selection techniques, when coupled with
optical follow-up, provide a reliable way to select clean samples
of clusters for cosmological analysis.  

\begin{figure}[htb]
 \begin{center}
  \includegraphics[scale=0.45]{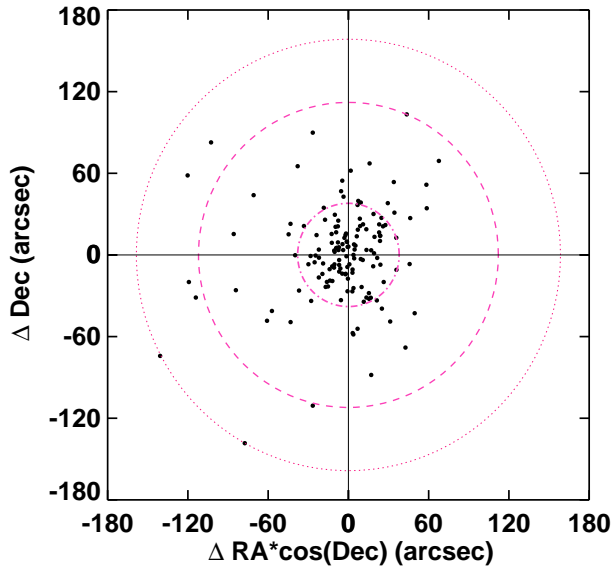}
  \caption{rBCG positions are plotted as offsets from SPT candidate
    positions for 146 systems with clearly defined rBCGs.  The magenta 
    concentric circles enclose 68\% (dash-dot line), 95\% (dashed line) and 99\% (dotted line) of the whole rBCG
    sample and have radii of 38.0\arcsec, 112.1\arcsec and
    158.6\arcsec.}
  \label{fig:BCG_offset}
  \end{center}
\end{figure}

\subsection{rBCG Offsets in SPT Clusters}
\label{sec:result:bcg}

The position of the rBCG in galaxy clusters is a property of interest
for both astrophysical and cosmological cluster studies, as it is a
possible indication of a cluster's dynamical state.  In relaxed
clusters, it is expected that dynamical friction will tend to drive
the most massive galaxies to the bottom of the cluster
potential well, which would coincide with the centroid of the X--ray and
SZ signatures.  On the other hand, in cases of merging systems one
would expect two different rBCGs, and one or both could appear well-separated from the X-ray or SZ centroid.
Several studies have shown a tight correlation between the X-ray
centroid and the rBCG position \citep{lin04b,haarsma10,mann12,stott12},
although \citet{fassbender11} provide evidence that at high redshift
the BCG distribution is less centrally peaked.  Here we examine the rBCG positions with respect to the centroid of
the SZ signal in the SPT cluster sample.

The position of each rBCG is listed in Table~\ref{tab:master list},
and the offsets from the SZ centroids in arcsec are plotted in
Figure~\ref{fig:BCG_offset}.  The rings correspond to different
fractions of the full population of clusters: 68\%, 95\%, and 99\%.
These rings have radii of 38.0\arcsec, 112.1\arcsec and 158.6\arcsec,
respectively.  The rBCG population is centrally concentrated with the
bulk of the SPT selected clusters having rBCGs lying within about 1$'$
of the candidate position.

\begin{figure}[htb]
 \begin{center}
   \includegraphics[scale=0.70]{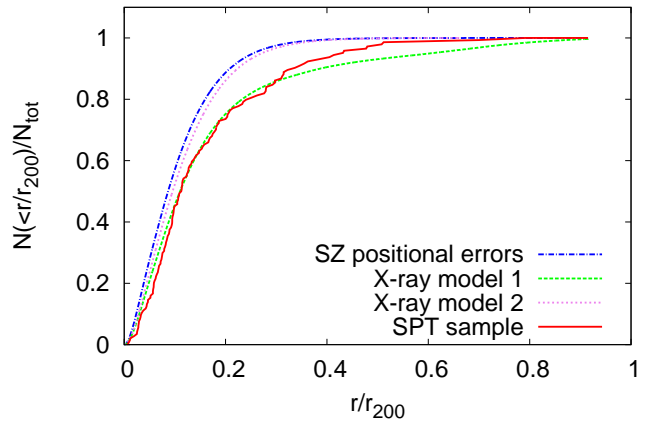}
   \caption{Normalized cumulative distribution of rBCG offsets 
     from SPT candidate positions as a function of
     $r/r_{200}$ for the SPT cluster sample (red solid line), the expected
     distribution given SPT positional uncertainties if all rBCGs were
     at exactly the center of the true SZ centroid (blue dash-dot line), and the
     expected distribution given SPT positional uncertainties if the
     underlying distribution of offsets matched those of two different X-ray
     selected cluster samples \citep[green dashed line;][]{lin04b} and \citep[magenta dotted line;][]{mann12}.  The KS
     probability that the observed rBCG distribution and the SPT
     positional error distribution are drawn from the same parent
     distribution is 0.09\%, but the observed distribution is statistically consistent with the
     distribution from the X-ray-selected sample convolved with the SPT positional uncertainty distribution.  There is no
     evidence in the rBCG offset distribution that SPT-selected
     clusters are more merger-rich than X-ray-selected clusters.}
   \label{fig:offset_distribution}
   \end{center}
\end{figure}

Given the broad redshift range of the cluster sample, the rBCG distribution in cluster coordinates $r/r_{200}$ is more physically interesting.
We use the cluster redshifts from this work and the SZ-derived masses
from R12 to calculate $r_{200}$ for each cluster.  The red line in
Figure~\ref{fig:offset_distribution} is the cumulative distribution of
the rBCGs as a function of $r/r_{200}$.  In this distribution, 68\% of
the rBCGs lies within 0.17$r_{200}$, 95\% within 0.43$r_{200}$ and 99\%
within 0.70$r_{200}$.

We check for any effects of mm-wave selection and redshift estimation
on the rBCG offset distribution by splitting the sample three ways: 1) clusters selected 
using 150~GHz data only vs.~clusters selected using multiband data; 
2) clusters with spectroscopic redshifts vs.~clusters with photometric
redshifts only; 3) clusters with secure photometric redshifts (Flag = 1) 
vs.~clusters with flagged redshifts (Flag $>$ 1, see \S\ref{sec:results} for 
details).  We see no evidence that the rBCG offsets (in units of
$r_{200}$) in these subsamples
are statistically different.  A Kolmogorov-Smirnov (KS) test results in  
probabilities of $84 \%$, $39 \%$, and $34 \%$ that these respective
subsamples are drawn from the same underlying distribution.

We investigate the importance of the SPT candidate positional
uncertainty by modeling the expected radial distribution in the case
where all rBCGs are located exactly at the cluster center.  The $1
\sigma$ SPT positional uncertainty for a cluster with a pressure
profile given by a spherical $\beta$ model with $\beta=1$ and scale
size $\theta_c$, detected by SPT at significance $\xi$, is given by
\begin{equation}
\Delta \theta = \sqrt{(\theta_\mathrm{beam}^2+(k \theta_{c})^2)}/\xi,
\label{eqn:posunc}
\end{equation}
where $\theta_\mathrm{beam}$ is the beam FWHM, and $k$ is a factor of
order unity (see \citealt{story11} for more details).  With this
information, we estimate the expected cumulative distribution of the
observed rBCG offsets, assuming a Gaussian with the appropriate width
for each cluster; this is equivalent to assuming the underlying rBCG
distribution is a delta function centered at zero offset with respect
to the true cluster SZ centroid.  Results are shown as the blue dotted
curve in Figure~\ref{fig:offset_distribution}.  It is clear that the
observed distribution of rBCG offsets is broader than that expected if
all rBCGs were located exactly at the cluster center.  We conduct
a Kolmogorov-Smirnov (KS) test to address the similarity of the two
distributions.  The hypothesis that the two distributions are drawn
from the same parent distribution has a probability of 0.09\%,
suggesting that the observed rBCG offset distribution cannot be easily
explained by the SPT positional uncertainties alone.

Because the SPT candidate positional uncertainties $\Delta \theta$ are
roughly the same, one can expect that our ability to measure the
underlying rBCG distribution will weaken as we push to higher redshift
where the cluster virial regions subtend smaller angles on the sky.
We test this by dividing the sample into four redshift bins with
similar numbers of members.  The KS tests confirm our expectations;
using redshift bins of 0.0-0.40, 0.40-0.54, 0.54-0.73 and $z>
0.73$ we find the probability that the observed and positional error
distributions are drawn from the same parent distribution is 0.11\%,
0.008\%, 1.97\% and 43.4\%, respectively.  Thus, with the current cluster
sample, we cannot detect any extent in the rBCG distribution beyond a
redshift $z\sim0.7$.  If we assume the underlying rBCG offset distribution is Gaussian, the KS test shows a maximum probability of 5.3\% for a Gaussian distributed width of 0.074$r_{200}$ with the probability of consistency dropping below 0.1\% for $\sigma > 0.08 r_{200}$.  Therefore, while the Gaussian is not a particularly good fit, the measured distribution strongly favors $\sigma < 0.08 r_{200}$.

We test whether our SZ-selected cluster sample exhibits similar rBCG
offsets to those seen in previous X-ray studies.  To do this, we adopt
the previously published BCG offset distribution from the X-ray
studies as the underlying BCG offset distribution for our sample and
then convolve this distribution with the SPT candidate positional
uncertainties.  If rBCGs in SZ-selected clusters are no different from
those in these previously studied samples, then we would expect the KS
test probability of consistency to be high.  We explore two samples:
X-ray model 1 
(\citealt{lin04b}; green dashed line in Figure~\ref{fig:offset_distribution})
and X-ray model 2
(\citealt{mann12}; magenta dotted line in Figure~\ref{fig:offset_distribution}).
The probability of
consistency between the SPT sample and X-ray model 1 is 41\%, and the
probability of consistency between the SPT sample and X-ray model 2 is
0.46\%.  We also examine another X-ray sample \citep{stott12} which produces a very similar result with our X-ray model 2 with the probability of consistency of 0.55\%.

It appears likely that the differences between the two previously
published X-ray samples can be explained in terms of differences in
the BCG selection.  The measured rBCG offset distribution presented
in this work agrees with the \citet{lin04b} sample, in which the BCGs were
defined as the brightest $K$-band galaxy projected within the virial
radius $\theta_{200}$ with spectroscopic redshift consistent with the
cluster redshift.  This BCG selection is very similar to the SPT rBCG
selection, with the main difference being that we do not have
spectroscopic redshifts for all rBCG candidates in the SPT sample.  The agreement between the SZ- and X-ray-selected samples in this case suggests that there are no strong differences between the
merger fractions in these two cluster samples.

The \citet{mann12} BCG sample, in contrast, was assembled using bluer
optical bands, which are more sensitive to the star formation history.
In addition, in cases where a second concentration of galaxies was
found within the projected virial region, the central galaxy of the
galaxy concentration coincident with the X-ray emission peak was
chosen as the BCG, regardless of whether it was brighter or not (Mann,
private communication).  This selection criteria would make it difficult to identify significantly offset BCGs, which would be more likely to be present in merging systems.  Similarly, the \citet{stott12} BCG sample was assembled using $i$-band data and a prior on the offset that excludes any offset greater than 500 kpc.  Such a prior would also bias the measured distribution against large offsets due to ongoing merger activity.

The rBCG offset distribution measured in the SPT
 SZ-selected sample of clusters does not provide any compelling
evidence that SZ-selected clusters differ in their merger rate as compared to X-ray-selected clusters.  It will be possible to test this more
precisely once we have the more accurate X-ray cluster centers with consistent rBCG selections.  Currently, the X-ray properties of only 15 SPT SZ-selected clusters have been published \citep{andersson11,benson11}; however, over 100 additional SPT selected clusters have been approved for observation in on-going programs with \emph{Chandra} and XMM-\emph{Newton}.   With those
data in hand we will be able to measure the rBCG offset distribution
over the full redshift range of SPT clusters, allowing us to probe for
evolution in the merger rates with redshift.

\section{Conclusions}
\label{sec:conclusion}

The SPT-SZ survey has produced an approximately mass-limited, redshift-independent sample 
of clusters.  Approximately 80\% of these clusters are newly discovered systems; the SPT survey has significantly increased the number of clusters discovered through the SZ effect and the number of massive clusters detected at high redshift.  In this paper, we present optical/NIR properties of 224 galaxy cluster
candidates selected from 720~deg$^2$ of the SPT survey that was
completed in 2008 and 2009.  The results
presented here constitute the subset of the survey in which the
optical/NIR follow-up is essentially complete.

With a dedicated pointed follow-up campaign using ground- and
space-based optical and NIR telescopes, we confirm 158 out of 224
SPT cluster candidates and measure their photometric redshifts.  We show that 18\% of the optically confirmed sample lies at $z > 0.8$, the median redshift is $z=0.57$, and the highest redshift cluster is at $z=1.35\pm0.07$.
We have undertaken a cross-comparison among three different cluster
redshift estimators 
to maximize the precision in the presented
photometric redshifts.  For each cluster, we combine the
redshift estimates from the three methods, accounting for the covariance
among the methods.  Using 57 clusters with spectroscopic redshifts, 
we calibrate the photometric redshifts and uncertainties and demonstrate that our combining 
procedure provides a characteristic final cluster redshift accuracy 
of $\Delta z/(1+z)=0.017$.

For the 66 candidates without photometric redshift measurements, we calculate
lower redshift limits.  These limits are set by 
the depth of the optical/NIR imaging and the band
combinations used.
For nine of these candidates there is evidence from NIR data 
that the cluster is a high
redshift system, and that we simply need deeper NIR data to measure a
photometric redshift.

Under the assumption that all 66 candidates
without photometric redshift measurements are noise fluctuations, we
estimate the purity of the SPT selected cluster sample as a function of
the SPT detection significance $\xi$.  Results are in good agreement
with expectations for sample purity, with no single unconfirmed system
above $\xi=6$, $>95\%$ purity above $\xi=5$, and $\sim70\%$ purity for
$\xi>4.5$.  By requiring an optical/NIR counterpart for each SPT
candidate, the purity in the final cluster sample approaches 100\% over
the full $\xi>4.5$ sample.  The purity of the SPT cluster sample
simplifies its cosmological interpretation.

Next, we examine the measured rBCG offset from the SZ candidate
positions to explore whether SZ-selected clusters exhibit similar
levels of ongoing merging as X-ray selected samples.  We show that
the characteristic offset between the rBCG and the candidate position
is $\sim$0.5$'$.   We examine the radial
distribution of rBCG offsets as a function of scaled cluster radius
$r/r_{200}$ and show that a model where we include scatter due to SPT positional uncertainties
assuming all BCGs are at cluster centers has only a 0.09\% chance of consistency with the observed
distribution.  That is, the observed distribution is broader than
would be expected from SPT positional uncertainties alone.  
If we assume the rBCG offset distribution is Gaussian, the observations rule out a Gaussian width of $\sigma > 0.08 r_{200}$, however, even with smaller width a Gaussian distribution is only marginally consistent with the data.  When comparing the SPT rBCG distribution with a X-ray selected cluster sample with a similar rBCG selection criteria \citep{lin04b}, the SPT and X-ray selected rBCG distributions are similar, suggesting that their merger rates are also similar.  Comparisons to other X-ray selected samples are complicated by differences in rBCG selection criteria.  For example, comparing to \citet{mann12}, which selects BCGs using bluer optical bands, we find a significantly less consistent rBCG distribution compared to SPT.  We conclude that SZ and X-ray selected cluster samples show consistent rBCG distributions, and note that BCG selection criteria can have a significant effect in such comparisons.

With the full 2500 \degs \ SZ survey completed in 2011, we are now working to complete the confirmations and redshift measurements of the full cluster candidate sample.  Scaling from this 720~deg$^2$ sample with effectively complete optical follow-up, we estimate that the full survey will produce $\sim$500 confirmed clusters, 
with approximately 100 of them at $z>0.8$.  This sample of clusters will enable an important next step in cluster cosmological studies as well as the first detailed glimpse of the high redshift tail of young, massive clusters.

\acknowledgments

The South Pole Telescope program is supported by the National Science
Foundation through grant ANT-0638937.  Partial support is also
provided by the NSF Physics Frontier Center grant PHY-0114422 to the
Kavli Institute of Cosmological Physics at the University of Chicago,
the Kavli Foundation, and the Gordon and Betty Moore Foundation.  The
Munich group acknowledges support from the Excellence Cluster Universe
and the DFG research program TR33 The Dark Universe.  Galaxy cluster
research at Harvard is supported by NSF grant AST-1009012, and
research at SAO is supported in part by NSF grants AST-1009649 and
MRI-0723073.  The McGill group acknowledges funding from the National
Sciences and Engineering Research Council of Canada, Canada Research
Chairs program, and the Canadian Institute for Advanced Research.

Optical imaging data from the Blanco 4~m at Cerro Tololo Interamerican
Observatories (programs 2005B-0043, 2009B-0400, 2010A-0441,
2010B-0598) and spectroscopic observations from VLT programs
086.A-0741 and 286.A-5021 and Gemini program GS-2009B-Q-16 were
included in this work. Additional data were obtained with the 6.5~m
Magellan Telescopes and the Swope telescope, which are located at the
Las Campanas Observatory in Chile.  This work is based in part on
observations made with the Spitzer Space Telescope (PIDs 60099,
70053), which is operated by the Jet Propulsion Laboratory, California
Institute of Technology under a contract with NASA. Support for this
work was provided by NASA through an award issued by JPL/Caltech.  The Digitized Sky Surveys were produced at the Space Telescope Science Institute under U.S. Government grant NAG W-2166. The images of these surveys are based on photographic data obtained using the Oschin Schmidt Telescope on Palomar Mountain and the UK Schmidt Telescope. The plates were processed into the present compressed digital form with the permission of these institutions.

{\it Facilities:}
\facility{Blanco (MOSAIC)},
\facility{Blanco (NEWFIRM)},
\facility{Gemini-S (GMOS)},
\facility{Magellan:Baade (IMACS)},
\facility{Magellan:Clay (LDSS3)},
\facility{South Pole Telescope},
\facility{\spitzer/IRAC},
\facility{Swope},
\facility{VLT:Antu (FORS2)},
\facility{\wise}

\bibliography{2008-2009_companion.bbl}

\newpage
\begin{appendix}

\section{Notable clusters}
\label{app:notable}

\paragraph{SPT-CL~J0337-6207}
This candidate is optically unconfirmed but has small NIR
blank-field probability in at least one data set
($\pblank = 0.5 \%$ in WISE data---also, $\pblank = 18.2 \%$ in
NEWFIRM data---see \S \ref{sec:method:nirrich} for details).

\paragraph{SPT-CL~J0428-6049}
This candidate is optically unconfirmed but has a high SPT
significance ($\xi = 5.1$) and small NIR blank-field probability in at
least one data set ($\pblank = 0.5 \%$ in WISE data, see \S
\ref{sec:method:nirrich} for details).  Though $\pblank = 69.0 \%$ in
NEWFIRM data, there is clear visual evidence of a NIR overdensity in
the NEWFIRM images, but at $\sim 40\arcsec$ from the SPT
position. Such a large offset is heavily disfavored by the fitting
procedure, such that the model that minimizes the overall $\chi^2$ for
the NEWFIRM data is effectively a blank field.  The position of the
WISE overdensity is consistent with the SPT position.

\paragraph{SPT-CL~J0458-5741}
This cluster is listed as optically unconfirmed, but it is also listed
in Table~2 of R12 as coincident with the low-redshift cluster ACO~3298
(at a separation of $77 \arcsec$).  We see a clear red-sequence
overdensity in our Magellan/IMACS data at $z \simeq 0.2$, centered on
the Abell cluster position.  The best-fit SZ core radius for this
candidate is $2.5 \arcmin$, which implies an SPT positional
uncertainty of $\sim 0.5 \arcmin$, in which case a $77 \arcsec$ offset
is only a $\sim 2 \sigma$ outlier.  However, visual inspection of a
lightly filtered SPT map shows that the SZ signal is coming from two
distinct components, one of which corresponds to the Abell cluster
position, and neither of which would have been significant enough to
be included in the R12 catalog on its own.  For this reason, we leave
the $\theta_c=2.5 \arcmin$ candidate, which blends the SZ signal from
the two individual components, as unconfirmed.

\paragraph{SPT-CL~J2002-5335}
This candidate is optically unconfirmed but has small NIR blank-field
probability in at least one data set ($\pblank = 7.5 \%$ in WISE
data---see \S \ref{sec:method:nirrich} for details).

\paragraph{SPT-CL~J2032-5627}
This cluster is listed in Table~2 of R12 as coincident with the $z =
0.06$ cluster ClG~2028.3-5637~/~ACO~3685 (at a separation of $115
\arcsec$ from the literature Abell cluster position) and as coincident
with the $z=0.14$ cluster RXC~J2032.1-5627 (at a separation of $87
\arcsec$ from the reported REFLEX cluster position).  However, from
our Magellan/IMACS imaging data, we estimate a red-sequence redshift
of $z = 0.31 \pm 0.02$, and, using the red-sequence measurements as a
criterion for cluster member selection, we have obtained spectra for
32 cluster members using GISMO and have measured a robust
spectroscopic redshift of $\zspec = 0.2840$.  Examination of the
REFLEX spectroscopic catalog \citep{guzzo09} reveals that their
spectroscopic observations yielded five galaxies near their reported
redshift of $\zspec = 0.1380$ but also six galaxies within $2 \%$ of
the value we derive from our GISMO observations ($\zspec = 0.2840$).
The value of $\zspec = 0.0608$ for ACO~3685 is from only one galaxy
(and, while reported in \citealt{struble99}, is originally from
\citealt{fetisova81}).  We conclude that there are two clear optical
overdensities at different redshifts along the line of sight to this
SZ/X-ray system, and that the literature redshift of $z=0.0608$ for
ACO~3685 is probably incorrect.  Because of the redshift dependence of
the SPT selection function (see, e.g., \citealt{vanderlinde10}), it is
likely that the bulk of the SZ signal is coming from the
higher-redshift cluster.  We have obtained XMM--Newton data on this
system, and the X-ray and SZ signals have very similar morphology,
indicating that the X-ray signal is also predominantly associated with
the higher-redshift system.  This makes it likely that the $z=0.2840$
system is a massive cluster and that the $z=0.1380$ system is a
low-mass interloper, possibly the cluster originally identified as
ClG~2028.3-5637~/~ACO~3685.  SPT-CL~J2032-5627 is discussed
further---including weak lensing data from Magellan/Megacam---in
\citet{high12}.

\paragraph{SPT-CL~J2035-5614}
This candidate is optically unconfirmed but has small NIR blank-field
probability in at least one data set ($\pblank = 0.1 \%$ in WISE
data---see \S \ref{sec:method:nirrich} for details).

\paragraph{SPT-CL~J2039-5723}
This candidate is optically unconfirmed but has a small SZ core radius
($0.5 \arcmin$) and small NIR blank-field probability in at least one
data set ($\pblank = 1.2 \%$ in WISE data and $8.7 \%$ in NEWFIRM
data---see \S \ref{sec:method:nirrich} for details).

\paragraph{SPT-CL~J2121-5546}
This candidate is optically unconfirmed but has small NIR blank-field
probability in at least one data set ($\pblank = 0.9 \%$ in WISE
data--also, $\pblank = 11.5 \%$ in NEWFIRM data---see \S
\ref{sec:method:nirrich} for details).

\paragraph{SPT-CL~J2136-5535}
This candidate is optically unconfirmed but has small NIR blank-field
probability in at least one data set ($\pblank = 5.2 \%$ in WISE
data---see \S \ref{sec:method:nirrich} for details).

\paragraph{SPT-CL~J2152-4629}
This candidate is optically unconfirmed but has a high SPT
significance ($\xi = 5.6$), a small SZ core radius ($0.25 \arcmin$),
and small NIR blank-field probability in at least one data set
($\pblank = 8.0 \%$ in WISE data; also $\pblank = 10.6 \%$ in
\spitzer/IRAC data and $20.0 \%$ in NEWFIRM data---see \S
\ref{sec:method:nirrich} for details).  This is the only candidate
with \spitzer/IRAC $\pblank < 20 \%$ for which we were not able to
estimate a redshift.

\paragraph{SPT-CL~J2343-5556}
This candidate is optically unconfirmed but has small NIR blank-field
probability in at least one data set ($\pblank = 5.6 \%$ in WISE data
and $21.0 \%$ in NEWFIRM data---see \S \ref{sec:method:nirrich} for
details).

\end{appendix}

\clearpage

\LongTables
\pagestyle{empty}
\begin{landscape}
\begin{center}
\def\arraystretch{1.2}
\tabletypesize{\scriptsize}
\begin{deluxetable}{ lrrc |  ccc  | ccc  | cc | c  }
\tablecaption{\label{tab:master list}All candidates above $\xi = 4.5$ in 720 deg$^2$ of the SPT-SZ survey.}
\tablehead{
\multicolumn{1}{l}{\bf{SPT ID}}  &
\multicolumn{2}{c}{\bf{POSITION}} &
\multicolumn{1}{c}{\bf{$\xi$}} &
\multicolumn{1}{c}{\bf{\zspec}\tablenotemark{a}} &
\multicolumn{1}{c}{\bf{\zcomb $\pm$ $\sigma_{z_\mathrm{comb}}$}} &
\multicolumn{1}{c}{\bf{Flag}\tablenotemark{b}} &
\multicolumn{3}{c}{\bf{NIR blank field probability \pblank (\%)}\tablenotemark{c}} &
\multicolumn{2}{c}{\bf{rBCG POSITION}} &
\multicolumn{1}{c}{\bf{Imaging Ref.}\tablenotemark{d}} \\
\colhead{} &
\colhead{RA (deg)} &
\colhead{Dec (deg)} &
\colhead{} &
\colhead{} &
\colhead{(or redshift lower limit)} &
\colhead{} &
\colhead{NEWFIRM} &
\colhead{\spitzer} &
\colhead{\wise} &
\colhead{RA (deg)} &
\colhead{Dec (deg)} &
\colhead{}\\
}

\startdata
SPT-CLJ0000-5748  &    0.2496  &  -57.8066  &   5.48  &  0.702  &    0.68 $\pm$  0.03  &  1  &  --  &  --  &  --  &    0.2503  &  -57.8093  &  1,2 \\
SPT-CLJ0201-6051  &   30.3933  &  -60.8592  &   4.83  &  --  &  $>$ 1.0  &  --  &  61.3  &  --  &  100.0  &  --  &  --  &  1 \\
SPT-CLJ0203-5651 \tablenotemark{1}  &   30.8309  &  -56.8612  &   4.98  &  --  &  $>$ 1.0  &  --  &  28.1  &  --  &   56.5  &  --  &  --  &  1 \\
SPT-CLJ0205-5829  &   31.4437  &  -58.4855  &  10.54  &  1.322  &    1.30 $\pm$  0.07  &  3  &   8.0  &   2.0  &    1.8  &   31.4510  &  -58.4803  &  1 \\
SPT-CLJ0205-6432  &   31.2786  &  -64.5461  &   6.02  &  0.744  &    0.75 $\pm$  0.03  &  1  &   6.8  &   0.5  &   17.3  &   31.3244  &  -64.5583  &  1 \\
SPT-CLJ0209-5452  &   32.3491  &  -54.8794  &   4.52  &  --  &    0.42 $\pm$  0.03  &  1  &  --  &  --  &  --  &   32.3494  &  -54.8720  &  1 \\
SPT-CLJ0211-5712  &   32.8232  &  -57.2157  &   4.77  &  --  &  $>$ 1.0  &  --  &  26.7  &  --  &   56.1  &  --  &  --  &  1 \\
SPT-CLJ0216-5730  &   34.1363  &  -57.5100  &   4.72  &  --  &  $>$ 1.0  &  --  &  43.4  &  --  &   80.6  &  --  &  --  &  1,2 \\
SPT-CLJ0216-6409  &   34.1723  &  -64.1562  &   5.54  &  --  &    0.64 $\pm$  0.03  &  1  &  --  &  --  &  --  &   34.1599  &  -64.1599  &  1 \\
SPT-CLJ0218-5826  &   34.6251  &  -58.4386  &   4.54  &  --  &    0.57 $\pm$  0.03  &  1  &  --  &  --  &  --  &   34.6267  &  -58.4421  &  1,2 \\
SPT-CLJ0221-6212  &   35.4382  &  -62.2044  &   4.71  &  --  &  $>$ 1.2  &  --  &  75.7  &  --  &   84.9  &  --  &  --  &  1 \\
SPT-CLJ0230-6028  &   37.6410  &  -60.4694  &   5.88  &  --  &    0.74 $\pm$  0.08  &  3  &  --  &  11.3  &    0.3  &   37.6354  &  -60.4628  &  3 \\
SPT-CLJ0233-5819  &   38.2561  &  -58.3269  &   6.64  &  0.663  &    0.76 $\pm$  0.07  &  3  &   4.2  &   0.0  &  --  &   38.2541  &  -58.3269  &  1 \\
SPT-CLJ0234-5831  &   38.6790  &  -58.5217  &  14.65  &  0.415  &    0.42 $\pm$  0.03  &  1  &  --  &  --  &  --  &   38.6762  &  -58.5236  &  1 \\
SPT-CLJ0239-6148  &   39.9120  &  -61.8032  &   4.67  &  --  &  $>$ 1.1  &  --  &  44.2  &  --  &   38.0  &  --  &  --  &  1,2 \\
SPT-CLJ0240-5946  &   40.1620  &  -59.7703  &   9.04  &  0.400  &    0.41 $\pm$  0.03  &  1  &  --  &  --  &  --  &   40.1599  &  -59.7635  &  3 \\
SPT-CLJ0240-5952  &   40.1982  &  -59.8785  &   4.65  &  --  &    0.62 $\pm$  0.03  &  3  &  --  &  --  &  --  &   40.2048  &  -59.8732  &  5 \\
SPT-CLJ0242-6039  &   40.6551  &  -60.6526  &   4.92  &  --  &  $>$ 1.5  &  --  &  --  &  51.5  &   64.9  &  --  &  --  &  2 \\
SPT-CLJ0243-5930  &   40.8616  &  -59.5132  &   7.42  &  --  &    0.65 $\pm$  0.03  &  1  &  --  &  --  &  --  &   40.8628  &  -59.5172  &  3 \\
SPT-CLJ0249-5658  &   42.4068  &  -56.9764  &   5.44  &  --  &    0.23 $\pm$  0.02  &  1  &  --  &  --  &  --  &   42.3918  &  -56.9870  &  1 \\
SPT-CLJ0253-6046  &   43.4605  &  -60.7744  &   4.83  &  --  &    0.44 $\pm$  0.02  &  1  &  --  &  --  &  --  &   43.4508  &  -60.7499  &  1 \\
SPT-CLJ0254-5857  &   43.5729  &  -58.9526  &  14.42  &  0.438  &    0.43 $\pm$  0.03  &  1  &  --  &  --  &  --  &   43.5365  &  -58.9718  &  3 \\
SPT-CLJ0254-6051  &   43.6015  &  -60.8643  &   6.71  &  --  &    0.44 $\pm$  0.02  &  1  &  --  &  --  &  --  &   43.5884  &  -60.8689  &  1 \\
SPT-CLJ0256-5617  &   44.1009  &  -56.2973  &   7.54  &  --  &    0.63 $\pm$  0.03  &  1  &  --  &  --  &  --  &   44.0880  &  -56.3031  &  3 \\
SPT-CLJ0257-5732  &   44.3516  &  -57.5423  &   5.40  &  0.434  &    0.42 $\pm$  0.02  &  1  &  --  &  --  &  --  &   44.3373  &  -57.5484  &  1 \\
SPT-CLJ0257-5842  &   44.3924  &  -58.7117  &   5.38  &  --  &    0.42 $\pm$  0.02  &  1  &  --  &  --  &  --  &   44.4374  &  -58.7045  &  1 \\
SPT-CLJ0257-6050  &   44.3354  &  -60.8450  &   4.76  &  --  &    0.48 $\pm$  0.03  &  1  &  --  &  --  &  --  &   44.3386  &  -60.8358  &  2 \\
SPT-CLJ0258-5756  &   44.5563  &  -57.9438  &   4.50  &  --  &  $>$ 1.0  &  --  &  --  &  --  &   17.4  &  --  &  --  &  1,2 \\
SPT-CLJ0300-6315  &   45.1430  &  -63.2643  &   4.88  &  --  &  $>$ 1.5  &  --  &  --  &  31.6  &   63.6  &  --  &  --  &  1 \\
SPT-CLJ0301-6456  &   45.4780  &  -64.9470  &   4.94  &  --  &    0.65 $\pm$  0.03  &  1  &  --  &  --  &  --  &   45.4809  &  -64.9492  &  1 \\
SPT-CLJ0307-6226  &   46.8335  &  -62.4336  &   8.32  &  --  &    0.61 $\pm$  0.03  &  1  &  --  &  --  &  --  &   46.8495  &  -62.4028  &  3 \\
SPT-CLJ0311-6354  &   47.8283  &  -63.9083  &   7.33  &  --  &    0.30 $\pm$  0.02  &  1  &  --  &  --  &  --  &   47.8229  &  -63.9157  &  1 \\
SPT-CLJ0313-5645  &   48.2604  &  -56.7554  &   4.82  &  --  &    0.63 $\pm$  0.03  &  1  &  --  &  --  &  --  &   48.2912  &  -56.7420  &  1 \\
SPT-CLJ0316-6059  &   49.2179  &  -60.9849  &   4.59  &  --  &  $>$ 1.5  &  --  &  --  &  26.4  &   26.9  &  --  &  --  &  1 \\
SPT-CLJ0317-5935  &   49.3208  &  -59.5856  &   5.91  &  0.469  &    0.47 $\pm$  0.02  &  1  &  --  &  --  &  --  &   49.3160  &  -59.5915  &  1 \\
SPT-CLJ0320-5800  &   50.0316  &  -58.0084  &   4.54  &  --  &  $>$ 1.0  &  --  &  --  &  --  &   38.1  &  --  &  --  &  1,2 \\
SPT-CLJ0324-6236  &   51.0530  &  -62.6018  &   8.59  &  --  &    0.74 $\pm$  0.03  &  1  &  --  &   0.3  &    0.0  &   51.0511  &  -62.5988  &  3 \\
SPT-CLJ0328-5541  &   52.1663  &  -55.6975  &   7.08  &  0.084  &    0.10 $\pm$  0.03  &  1  &  --  &  --  &  --  &   52.1496  &  -55.7124  &  1 \\
SPT-CLJ0333-5842  &   53.3195  &  -58.7019  &   4.54  &  --  &    0.49 $\pm$  0.04  &  3$^*$  &  --  &  --  &  --  &   53.3322  &  -58.7060  &  5 \\
SPT-CLJ0337-6207  &   54.4720  &  -62.1176  &   4.88  &  --  &  $>$ 1.3  &  --  &  18.2  &  --  &    0.5  &  --  &  --  &  1 \\
SPT-CLJ0337-6300  &   54.4685  &  -63.0098  &   5.29  &  --  &    0.46 $\pm$  0.03  &  1  &  --  &  --  &  --  &   54.4744  &  -63.0155  &  1 \\
SPT-CLJ0341-5731  &   55.3979  &  -57.5233  &   5.35  &  --  &    0.64 $\pm$  0.02  &  1  &  --  &  --  &  --  &   55.3955  &  -57.5244  &  1 \\
SPT-CLJ0341-6143  &   55.3485  &  -61.7192  &   5.60  &  --  &    0.63 $\pm$  0.03  &  1  &  --  &  --  &  --  &   55.3488  &  -61.7208  &  1 \\
SPT-CLJ0343-5518  &   55.7634  &  -55.3049  &   5.98  &  --  &    0.49 $\pm$  0.02  &  1  &  --  &  --  &  --  &   55.7581  &  -55.3111  &  1 \\
SPT-CLJ0344-5452  &   56.0926  &  -54.8726  &   5.41  &  --  &    1.01 $\pm$  0.07  &  3  &  --  &   8.0  &   17.6  &  --  &  --  &  1 \\
SPT-CLJ0344-5518  &   56.2101  &  -55.3036  &   5.02  &  --  &    0.36 $\pm$  0.02  &  1  &  --  &  --  &  --  &   56.1816  &  -55.3179  &  1 \\
SPT-CLJ0345-6419  &   56.2518  &  -64.3326  &   5.57  &  --  &    0.93 $\pm$  0.07  &  3  &   3.4  &   0.3  &    0.1  &   56.2518  &  -64.3343  &  1 \\
SPT-CLJ0346-5839  &   56.5746  &  -58.6535  &   4.96  &  --  &    0.74 $\pm$  0.07  &  3  &  --  &   0.4  &    0.4  &   56.5754  &  -58.6532  &  1 \\
SPT-CLJ0351-5636  &   57.9312  &  -56.6099  &   4.65  &  --  &    0.38 $\pm$  0.03  &  1  &  --  &  --  &  --  &   57.9446  &  -56.6349  &  1 \\
SPT-CLJ0351-5944  &   57.8654  &  -59.7457  &   4.61  &  --  &  $>$ 1.0  &  --  &  20.4  &  --  &   60.4  &  --  &  --  &  2 \\
SPT-CLJ0352-5647  &   58.2366  &  -56.7992  &   7.11  &  --  &    0.66 $\pm$  0.03  &  1  &  --  &  --  &  --  &   58.2759  &  -56.7608  &  3 \\
SPT-CLJ0354-5904  &   58.5611  &  -59.0741  &   6.49  &  --  &    0.41 $\pm$  0.03  &  1  &  --  &  --  &  --  &   58.6166  &  -59.0971  &  1 \\
SPT-CLJ0354-6032 \tablenotemark{2}  &   58.6744  &  -60.5386  &   4.57  &  --  &    1.06 $\pm$  0.07  &  3  &  --  &   8.1  &    1.9  &   58.6604  &  -60.5462  &  2 \\
SPT-CLJ0402-6129  &   60.7066  &  -61.4988  &   4.83  &  --  &    0.53 $\pm$  0.04  &  3  &  --  &  --  &  --  &   60.7213  &  -61.4973  &  5 \\
SPT-CLJ0403-5534  &   60.9479  &  -55.5829  &   4.88  &  --  &  $>$ 1.5  &  --  &  --  &  71.1  &   60.5  &  --  &  --  &  1 \\
SPT-CLJ0403-5719  &   60.9670  &  -57.3241  &   5.75  &  --  &    0.43 $\pm$  0.02  &  1  &  --  &  --  &  --  &   60.9679  &  -57.3285  &  1 \\
SPT-CLJ0404-6510  &   61.0556  &  -65.1817  &   4.75  &  --  &    0.15 $\pm$  0.02  &  1  &  --  &  --  &  --  &   61.0934  &  -65.1703  &  1 \\
SPT-CLJ0406-5455  &   61.6922  &  -54.9205  &   5.82  &  --  &    0.73 $\pm$  0.03  &  1  &  --  &   3.3  &   19.6  &   61.6857  &  -54.9257  &  1 \\
SPT-CLJ0410-5454  &   62.6154  &  -54.9016  &   5.06  &  --  &  $>$ 1.0  &  --  &  76.8  &  --  &   35.0  &  --  &  --  &  1 \\
SPT-CLJ0410-6343  &   62.5158  &  -63.7285  &   5.79  &  --  &    0.49 $\pm$  0.02  &  1  &  --  &  --  &  --  &   62.5207  &  -63.7311  &  1 \\
SPT-CLJ0411-5751  &   62.8433  &  -57.8636  &   5.16  &  --  &    0.75 $\pm$  0.02  &  1  &  --  &  --  &   25.3  &   62.8174  &  -57.8517  &  1 \\
SPT-CLJ0411-6340  &   62.8597  &  -63.6810  &   6.41  &  --  &    0.14 $\pm$  0.02  &  1  &  --  &  --  &  --  &   62.8676  &  -63.6853  &  1 \\
SPT-CLJ0412-5743  &   63.0245  &  -57.7202  &   5.29  &  --  &    0.38 $\pm$  0.03  &  1  &  --  &  --  &  --  &   63.0442  &  -57.7383  &  1 \\
SPT-CLJ0416-6359  &   64.1618  &  -63.9964  &   6.06  &  --  &    0.30 $\pm$  0.02  &  1  &  --  &  --  &  --  &   64.1735  &  -64.0060  &  1 \\
SPT-CLJ0423-5506  &   65.8153  &  -55.1036  &   4.51  &  --  &    0.21 $\pm$  0.04  &  3  &  --  &  --  &  --  &   65.8108  &  -55.1143  &  5 \\
SPT-CLJ0423-6143  &   65.9366  &  -61.7183  &   4.65  &  --  &    0.71 $\pm$  0.04  &  3  &  --  &  --  &   13.9  &   65.9323  &  -61.7293  &  5 \\
SPT-CLJ0426-5455  &   66.5205  &  -54.9201  &   8.86  &  --  &    0.63 $\pm$  0.03  &  1  &  --  &  --  &  --  &   66.5171  &  -54.9253  &  3 \\
SPT-CLJ0428-6049  &   67.0291  &  -60.8302  &   5.06  &  --  &  $>$ 1.1  &  --  &  69.0  &  --  &    0.5  &  --  &  --  &  1 \\
SPT-CLJ0430-6251 \tablenotemark{3}  &   67.7086  &  -62.8536  &   5.20  &  --  &    0.38 $\pm$  0.04  &  4  &  --  &  --  &  --  &  --  &  --  &  1 \\
SPT-CLJ0431-6126  &   67.8393  &  -61.4438  &   6.40  &  0.058  &    0.08 $\pm$  0.02  &  1  &  --  &  --  &  --  &   67.8053  &  -61.4533  &  1 \\
SPT-CLJ0433-5630  &   68.2522  &  -56.5038  &   5.35  &  0.692  &    0.65 $\pm$  0.03  &  1  &  --  &  --  &  --  &   68.2545  &  -56.5190  &  1 \\
SPT-CLJ0441-5859  &   70.4411  &  -58.9931  &   4.54  &  --  &  $>$ 1.1  &  --  &  --  &  --  &   27.7  &  --  &  --  &  5 \\
SPT-CLJ0444-5603 \tablenotemark{4}  &   71.1130  &  -56.0566  &   5.30  &  --  &    0.98 $\pm$  0.07  &  3  &   0.4  &   0.0  &    0.0  &   71.1077  &  -56.0556  &  1 \\
SPT-CLJ0446-5849  &   71.5160  &  -58.8226  &   7.44  &  --  &    1.16 $\pm$  0.07  &  3  &  --  &   0.9  &    5.6  &   71.5138  &  -58.8247  &  1 \\
SPT-CLJ0452-5945  &   73.1282  &  -59.7622  &   4.50  &  --  &  $>$ 0.7  &  --  &  --  &  --  &   39.4  &  --  &  --  &  5 \\
SPT-CLJ0456-5623  &   74.1745  &  -56.3868  &   4.76  &  --  &    0.66 $\pm$  0.03  &  1  &  --  &  --  &  --  &  --  &  --  &  2 \\
SPT-CLJ0456-6141  &   74.1496  &  -61.6840  &   4.84  &  --  &    0.41 $\pm$  0.03  &  3$^*$  &  --  &  --  &  --  &   74.1361  &  -61.6902  &  5 \\
SPT-CLJ0458-5741  &   74.6021  &  -57.6952  &   4.91  &  --  &  $>$ 1.0  &  --  &  --  &  --  &   52.5  &  --  &  --  &  2 \\
SPT-CLJ0502-6113  &   75.5400  &  -61.2315  &   5.09  &  --  &    0.66 $\pm$  0.03  &  1  &  --  &  --  &  --  &   75.5630  &  -61.2314  &  1 \\
SPT-CLJ0509-5342  &   77.3360  &  -53.7046  &   6.61  &  0.461  &    0.43 $\pm$  0.02  &  1  &  --  &  --  &  --  &   77.3392  &  -53.7036  &  1,2 \\
SPT-CLJ0511-5154  &   77.9202  &  -51.9044  &   5.63  &  0.645  &    0.63 $\pm$  0.03  &  1  &  --  &  --  &  --  &  --  &  --  &  1$^{*}$ \\
SPT-CLJ0514-5118  &   78.6859  &  -51.3100  &   4.82  &  --  &  $>$ 1.2  &  --  &  28.2  &  --  &   49.7  &  --  &  --  &  1 \\
SPT-CLJ0516-5430  &   79.1480  &  -54.5062  &   9.42  &  0.295  &    0.31 $\pm$  0.02  &  1  &  --  &  --  &  --  &   79.1557  &  -54.5007  &  1,2 \\
SPT-CLJ0521-5104  &   80.2983  &  -51.0812  &   5.45  &  0.675  &    0.64 $\pm$  0.02  &  1  &  --  &  --  &  --  &   80.3106  &  -51.0718  &  1$^{*}$ \\
SPT-CLJ0522-5026  &   80.5190  &  -50.4409  &   4.87  &  --  &    0.51 $\pm$  0.03  &  1  &  --  &  --  &  --  &   80.5000  &  -50.4696  &  1$^{*}$ \\
SPT-CLJ0527-5928  &   81.8111  &  -59.4833  &   4.71  &  --  &  $>$ 0.9  &  --  &  --  &  --  &   25.5  &  --  &  --  &  2 \\
SPT-CLJ0528-5300  &   82.0173  &  -53.0001  &   5.45  &  0.768  &    0.77 $\pm$  0.03  &  1  &  --  &   0.0  &    0.0  &   82.0221  &  -52.9982  &  1$^{*}$ \\
SPT-CLJ0529-5238  &   82.2923  &  -52.6417  &   4.52  &  --  &  $>$ 1.1  &  --  &  --  &  --  &   81.8  &  --  &  --  &  1$^{*}$ \\
SPT-CLJ0532-5647  &   83.1586  &  -56.7893  &   4.51  &  --  &  $>$ 0.9  &  --  &  --  &  --  &   39.0  &  --  &  --  &  2 \\
SPT-CLJ0533-5005  &   83.3984  &  -50.0918  &   5.59  &  0.881  &    0.81 $\pm$  0.03  &  1  &  --  &   0.0  &   18.4  &   83.4144  &  -50.0845  &  2 \\
SPT-CLJ0534-5937  &   83.6018  &  -59.6289  &   4.57  &  0.576  &    0.57 $\pm$  0.02  &  1  &  --  &  --  &  --  &   83.6255  &  -59.6152  &  2 \\
SPT-CLJ0537-5549  &   84.2578  &  -55.8268  &   4.55  &  --  &  $>$ 1.1  &  --  &  --  &  --  &   28.4  &  --  &  --  &  2 \\
SPT-CLJ0538-5657  &   84.5865  &  -56.9530  &   4.63  &  --  &  $>$ 1.5  &  --  &  --  &  21.8  &   25.1  &  --  &  --  &  2 \\
SPT-CLJ0539-5744  &   84.9998  &  -57.7432  &   5.12  &  --  &    0.76 $\pm$  0.03  &  1  &  --  &   0.0  &    0.0  &   84.9950  &  -57.7424  &  2 \\
SPT-CLJ0546-5345  &   86.6541  &  -53.7615  &   7.69  &  1.066  &    1.04 $\pm$  0.07  &  3  &  --  &   0.0  &    0.0  &   86.6569  &  -53.7587  &  1$^{*}$ \\
SPT-CLJ0551-5709  &   87.9016  &  -57.1565  &   6.13  &  0.423  &    0.43 $\pm$  0.02  &  1  &  --  &  --  &  --  &   87.8981  &  -57.1414  &  2 \\
SPT-CLJ0556-5403  &   89.2016  &  -54.0630  &   4.83  &  --  &    0.93 $\pm$  0.04  &  4  &  17.1  &  --  &    0.0  &   89.2018  &  -54.0582  &  2 \\
SPT-CLJ0559-5249  &   89.9245  &  -52.8265  &   9.28  &  0.609  &    0.63 $\pm$  0.02  &  1  &  --  &  --  &  --  &   89.9301  &  -52.8242  &  2 \\
SPT-CLJ2002-5335  &  300.5113  &  -53.5913  &   4.53  &  --  &  $>$ 1.0  &  --  &  --  &  --  &    7.5  &  --  &  --  &  1 \\
SPT-CLJ2005-5635  &  301.3385  &  -56.5902  &   4.68  &  --  &  $>$ 0.6  &  --  &  --  &  --  &   71.3  &  --  &  --  &  1 \\
SPT-CLJ2006-5325  &  301.6620  &  -53.4286  &   5.06  &  --  &  $>$ 1.5  &  --  &  66.8  &  52.3  &   71.8  &  --  &  --  &  1 \\
SPT-CLJ2007-4906  &  301.9663  &  -49.1105  &   4.50  &  --  &    1.25 $\pm$  0.07  &  3  &  --  &   0.9  &    5.9  &  301.9692  &  -49.1085  &  1 \\
SPT-CLJ2009-5756  &  302.4261  &  -57.9480  &   4.68  &  --  &    0.63 $\pm$  0.03  &  1  &  --  &  --  &  --  &  --  &  --  &  1 \\
SPT-CLJ2011-5228 \tablenotemark{5}  &  302.7810  &  -52.4734  &   4.55  &  --  &    0.96 $\pm$  0.04  &  1  &  --  &  --  &   51.6  &  302.7814  &  -52.4709  &  1 \\
SPT-CLJ2011-5725 \tablenotemark{6}  &  302.8526  &  -57.4214  &   5.43  &  0.279  &    0.28 $\pm$  0.03  &  1  &  --  &  --  &  --  &  302.8624  &  -57.4197  &  1 \\
SPT-CLJ2012-5342  &  303.0822  &  -53.7137  &   4.65  &  --  &  $>$ 0.7  &  --  &  --  &  --  &   11.0  &  --  &  --  &  1 \\
SPT-CLJ2012-5649  &  303.1132  &  -56.8308  &   5.99  &  0.055  &    0.07 $\pm$  0.02  &  1  &  --  &  --  &  --  &  303.1142  &  -56.8270  &  2 \\
SPT-CLJ2013-5432  &  303.4968  &  -54.5445  &   4.75  &  --  &  $>$ 1.0  &  --  &  49.7  &  --  &   63.4  &  --  &  --  &  1 \\
SPT-CLJ2015-5504  &  303.9884  &  -55.0715  &   4.64  &  --  &  $>$ 0.6  &  --  &  87.3  &  --  &   67.0  &  --  &  --  &  1 \\
SPT-CLJ2016-4954  &  304.0181  &  -49.9122  &   5.01  &  --  &    0.26 $\pm$  0.02  &  1  &  --  &  --  &  --  &  304.0067  &  -49.9067  &  1 \\
SPT-CLJ2017-6258  &  304.4827  &  -62.9763  &   6.45  &  --  &    0.57 $\pm$  0.03  &  1  &  --  &  --  &  --  &  304.4730  &  -62.9950  &  3 \\
SPT-CLJ2018-4528  &  304.6076  &  -45.4807  &   4.64  &  --  &    0.40 $\pm$  0.03  &  1  &  --  &  --  &  --  &  304.6164  &  -45.4761  &  1 \\
SPT-CLJ2019-5642  &  304.7703  &  -56.7079  &   5.25  &  --  &    0.15 $\pm$  0.03  &  1  &  --  &  --  &  --  &  304.8137  &  -56.7122  &  2 \\
SPT-CLJ2020-4646  &  305.1936  &  -46.7702  &   5.09  &  --  &    0.17 $\pm$  0.02  &  1  &  --  &  --  &  --  &  305.1973  &  -46.7748  &  1 \\
SPT-CLJ2020-6314  &  305.0301  &  -63.2413  &   5.37  &  --  &    0.58 $\pm$  0.02  &  1  &  --  &  --  &  --  &  305.0350  &  -63.2471  &  2 \\
SPT-CLJ2021-5256  &  305.4690  &  -52.9439  &   5.31  &  --  &    0.11 $\pm$  0.02  &  1  &  --  &  --  &  --  &  305.4725  &  -52.9509  &  1 \\
SPT-CLJ2022-6323  &  305.5235  &  -63.3973  &   6.58  &  0.383  &    0.41 $\pm$  0.02  &  1  &  --  &  --  &  --  &  305.5410  &  -63.3971  &  2,4 \\
SPT-CLJ2023-5535  &  305.8377  &  -55.5903  &  13.41  &  0.232  &    0.22 $\pm$  0.02  &  1  &  --  &  --  &  --  &  305.9069  &  -55.5697  &  2,3 \\
SPT-CLJ2025-5117  &  306.4836  &  -51.2904  &   9.48  &  --  &    0.20 $\pm$  0.02  &  1  &  --  &  --  &  --  &  306.4822  &  -51.2744  &  1 \\
SPT-CLJ2026-4513  &  306.6140  &  -45.2256  &   5.53  &  --  &    0.71 $\pm$  0.03  &  1  &  --  &   2.1  &   18.8  &  306.6180  &  -45.2338  &  1 \\
SPT-CLJ2030-5638  &  307.7067  &  -56.6352  &   5.47  &  --  &    0.39 $\pm$  0.03  &  1  &  --  &  --  &  --  &  307.6886  &  -56.6322  &  2,4 \\
SPT-CLJ2032-5627  &  308.0800  &  -56.4557  &   8.14  &  0.284  &    0.33 $\pm$  0.02  &  1  &  --  &  --  &  --  &  308.0586  &  -56.4368  &  2 \\
SPT-CLJ2034-5936  &  308.5408  &  -59.6007  &   8.57  &  --  &    0.92 $\pm$  0.07  &  3  &  --  &   0.2  &   25.9  &  308.5414  &  -59.6034  &  2 \\
SPT-CLJ2035-5251  &  308.8026  &  -52.8527  &   9.99  &  --  &    0.47 $\pm$  0.02  &  1  &  --  &  --  &  --  &  --  &  --  &  1 \\
SPT-CLJ2035-5614  &  308.9023  &  -56.2407  &   4.55  &  --  &  $>$ 1.0  &  --  &  --  &  --  &    0.1  &  --  &  --  &  1 \\
SPT-CLJ2039-5723  &  309.8246  &  -57.3871  &   4.69  &  --  &  $>$ 1.2  &  --  &   9.2  &  --  &    1.2  &  --  &  --  &  1,2 \\
SPT-CLJ2040-4451  &  310.2468  &  -44.8599  &   6.28  &  --  &    1.35 $\pm$  0.07  &  3  &  29.9  &   4.6  &    3.6  &  310.2384  &  -44.8593  &  1 \\
SPT-CLJ2040-5230  &  310.1255  &  -52.5052  &   4.70  &  --  &  $>$ 1.0  &  --  &  44.4  &  --  &   20.2  &  --  &  --  &  1 \\
SPT-CLJ2040-5342  &  310.2195  &  -53.7122  &   5.88  &  --  &    0.57 $\pm$  0.04  &  1  &  --  &  --  &  --  &  --  &  --  &  1 \\
SPT-CLJ2040-5725  &  310.0631  &  -57.4287  &   6.38  &  0.930  &    0.91 $\pm$  0.07  &  3  &  --  &   0.9  &   15.0  &  310.0552  &  -57.4209  &  1,2 \\
SPT-CLJ2043-5035  &  310.8285  &  -50.5929  &   7.81  &  0.723  &    0.77 $\pm$  0.03  &  1  &  --  &   0.6  &    0.4  &  --  &  --  &  1 \\
SPT-CLJ2043-5614  &  310.7906  &  -56.2351  &   4.72  &  --  &    0.69 $\pm$  0.03  &  1  &  --  &  --  &  --  &  310.7788  &  -56.2390  &  1 \\
SPT-CLJ2045-6026  &  311.3649  &  -60.4469  &   4.77  &  --  &  $>$ 0.5  &  --  &  86.3  &  --  &   94.6  &  --  &  --  &  1 \\
SPT-CLJ2046-4542  &  311.5620  &  -45.7111  &   4.54  &  --  &  $>$ 1.0  &  --  &  --  &  --  &   65.3  &  --  &  --  &  1 \\
SPT-CLJ2048-4524  &  312.2268  &  -45.4150  &   4.56  &  --  &  $>$ 1.0  &  --  &  --  &  --  &   96.6  &  --  &  --  &  1 \\
SPT-CLJ2051-6256  &  312.8027  &  -62.9348  &   5.17  &  --  &    0.47 $\pm$  0.02  &  1  &  --  &  --  &  --  &  312.8230  &  -62.9407  &  1 \\
SPT-CLJ2055-5456  &  313.9941  &  -54.9366  &   6.61  &  --  &    0.11 $\pm$  0.02  &  1  &  --  &  --  &  --  &  313.9838  &  -54.9273  &  2 \\
SPT-CLJ2056-5106  &  314.0723  &  -51.1163  &   4.70  &  --  &  $>$ 1.0  &  --  &  43.0  &  --  &   45.4  &  --  &  --  &  1 \\
SPT-CLJ2056-5459  &  314.2199  &  -54.9892  &   6.05  &  0.718  &    0.84 $\pm$  0.07  &  3  &  --  &   0.3  &    9.3  &  314.2232  &  -54.9858  &  2 \\
SPT-CLJ2057-5251  &  314.4105  &  -52.8567  &   4.52  &  --  &  $>$ 1.5  &  --  &  --  &  82.1  &   65.8  &  --  &  --  &  1 \\
SPT-CLJ2058-5608  &  314.5893  &  -56.1454  &   5.02  &  0.606  &    0.59 $\pm$  0.02  &  1  &  --  &  --  &  --  &  314.5930  &  -56.1464  &  1 \\
SPT-CLJ2059-5018  &  314.9324  &  -50.3049  &   4.79  &  --  &    0.39 $\pm$  0.02  &  1  &  --  &  --  &  --  &  314.9220  &  -50.3029  &  1 \\
SPT-CLJ2100-4548  &  315.0936  &  -45.8057  &   4.84  &  0.712  &    0.74 $\pm$  0.03  &  1  &  --  &   1.2  &    6.9  &  --  &  --  &  1 \\
SPT-CLJ2100-5708  &  315.1502  &  -57.1347  &   5.11  &  --  &    0.59 $\pm$  0.03  &  1  &  --  &  --  &  --  &  315.1470  &  -57.1385  &  1 \\
SPT-CLJ2101-5542  &  315.3106  &  -55.7027  &   5.04  &  --  &    0.22 $\pm$  0.02  &  1  &  --  &  --  &  --  &  315.3040  &  -55.6940  &  1 \\
SPT-CLJ2101-6123  &  315.4594  &  -61.3972  &   5.28  &  --  &    0.60 $\pm$  0.03  &  1  &  --  &  --  &  --  &  315.4326  &  -61.4047  &  1 \\
SPT-CLJ2103-5411  &  315.7687  &  -54.1951  &   4.88  &  --  &    0.46 $\pm$  0.02  &  1  &  --  &  --  &  --  &  315.7792  &  -54.1945  &  1 \\
SPT-CLJ2104-5224  &  316.2283  &  -52.4044  &   5.32  &  0.799  &    0.81 $\pm$  0.03  &  1  &  --  &  65.7  &    3.0  &  316.2120  &  -52.4079  &  1 \\
SPT-CLJ2106-5820  &  316.5144  &  -58.3459  &   4.81  &  --  &  $>$ 1.0  &  --  &  79.4  &  --  &   60.6  &  --  &  --  &  1 \\
SPT-CLJ2106-5844  &  316.5210  &  -58.7448  &  22.08  &  1.132  &    1.20 $\pm$  0.07  &  3  &  --  &   0.0  &    0.1  &  316.5194  &  -58.7412  &  2 \\
SPT-CLJ2106-6019  &  316.6642  &  -60.3299  &   4.98  &  --  &    0.97 $\pm$  0.03  &  1  &   2.5  &  --  &    1.0  &  316.6449  &  -60.3385  &  1 \\
SPT-CLJ2106-6303  &  316.6596  &  -63.0510  &   4.90  &  --  &  $>$ 1.0  &  --  &  19.6  &  --  &   16.7  &  --  &  --  &  1 \\
SPT-CLJ2109-4626  &  317.4516  &  -46.4370  &   5.51  &  --  &    0.98 $\pm$  0.09  &  3  &  --  &   1.1  &    0.6  &  317.4557  &  -46.4376  &  1 \\
SPT-CLJ2109-5040  &  317.3820  &  -50.6773  &   5.17  &  --  &    0.47 $\pm$  0.03  &  1  &  --  &  --  &  --  &  317.4016  &  -50.6815  &  1 \\
SPT-CLJ2110-5244  &  317.5502  &  -52.7486  &   6.22  &  --  &    0.61 $\pm$  0.02  &  1  &  --  &  --  &  --  &  317.5520  &  -52.7496  &  1 \\
SPT-CLJ2111-5338  &  317.9217  &  -53.6496  &   5.65  &  --  &    0.43 $\pm$  0.03  &  1  &  --  &  --  &  --  &  317.9357  &  -53.6477  &  1 \\
SPT-CLJ2115-4659  &  318.7995  &  -46.9862  &   5.60  &  --  &    0.34 $\pm$  0.02  &  1  &  --  &  --  &  --  &  318.8064  &  -46.9797  &  1 \\
SPT-CLJ2118-5055  &  319.7291  &  -50.9329  &   5.62  &  0.625  &    0.63 $\pm$  0.03  &  1  &  --  &  --  &  --  &  --  &  --  &  1 \\
SPT-CLJ2119-6230  &  319.8846  &  -62.5096  &   4.55  &  --  &    0.72 $\pm$  0.03  &  1  &  --  &  --  &  --  &  319.8765  &  -62.5106  &  1 \\
SPT-CLJ2120-4728 \tablenotemark{7}  &  320.1594  &  -47.4776  &   5.98  &  --  &    0.99 $\pm$  0.07  &  3  &   8.7  &   0.5  &    2.1  &  320.1638  &  -47.4750  &  1 \\
SPT-CLJ2121-5546  &  320.2715  &  -55.7780  &   4.79  &  --  &  $>$ 0.8  &  --  &  11.5  &  --  &    0.9  &  --  &  --  &  2 \\
SPT-CLJ2121-6335  &  320.4269  &  -63.5843  &   5.43  &  --  &    0.23 $\pm$  0.02  &  1  &  --  &  --  &  --  &  320.4303  &  -63.5973  &  1 \\
SPT-CLJ2124-6124  &  321.1488  &  -61.4141  &   8.21  &  0.435  &    0.44 $\pm$  0.02  &  1  &  --  &  --  &  --  &  321.1577  &  -61.4077  &  1 \\
SPT-CLJ2125-6113  &  321.2902  &  -61.2292  &   4.74  &  --  &  $>$ 1.5  &  --  &  --  &  91.8  &   17.2  &  --  &  --  &  1,2 \\
SPT-CLJ2127-6443  &  321.9939  &  -64.7288  &   4.54  &  --  &  $>$ 1.0  &  --  &  --  &  --  &   80.2  &  --  &  --  &  1 \\
SPT-CLJ2130-4737  &  322.6622  &  -47.6257  &   4.83  &  --  &  $>$ 1.5  &  --  &  76.9  &  22.4  &   68.5  &  --  &  --  &  1 \\
SPT-CLJ2130-6458  &  322.7285  &  -64.9764  &   7.57  &  0.316  &    0.36 $\pm$  0.02  &  1  &  --  &  --  &  --  &  322.7343  &  -64.9779  &  1,2 \\
SPT-CLJ2131-5003  &  322.9717  &  -50.0647  &   4.83  &  --  &    0.45 $\pm$  0.02  &  1  &  --  &  --  &  --  &  322.9637  &  -50.0624  &  1 \\
SPT-CLJ2133-5411  &  323.2978  &  -54.1845  &   4.58  &  --  &  $>$ 1.5  &  --  &  --  &  48.7  &   32.0  &  --  &  --  &  1 \\
SPT-CLJ2135-5452  &  323.9060  &  -54.8773  &   4.61  &  --  &  $>$ 1.0  &  --  &  --  &  --  &   53.4  &  --  &  --  &  1 \\
SPT-CLJ2135-5726  &  323.9158  &  -57.4415  &  10.43  &  0.427  &    0.46 $\pm$  0.02  &  1  &  --  &  --  &  --  &  323.9059  &  -57.4418  &  2,4 \\
SPT-CLJ2136-4704  &  324.1175  &  -47.0803  &   6.17  &  0.425  &    0.43 $\pm$  0.03  &  1  &  --  &  --  &  --  &  324.1640  &  -47.0716  &  1 \\
SPT-CLJ2136-5519  &  324.2392  &  -55.3215  &   4.65  &  --  &  $>$ 1.5  &  --  &  --  &  35.2  &   62.1  &  --  &  --  &  1 \\
SPT-CLJ2136-5535 \tablenotemark{8}  &  324.0898  &  -55.5853  &   4.58  &  --  &  $>$ 1.2  &  --  &  --  &  --  &    5.2  &  --  &  --  &  1 \\
SPT-CLJ2136-5723  &  324.1209  &  -57.3923  &   4.55  &  --  &  $>$ 1.0  &  --  &  --  &  --  &   40.0  &  --  &  --  &  2,4 \\
SPT-CLJ2136-6307  &  324.2334  &  -63.1233  &   6.25  &  0.926  &    1.00 $\pm$  0.07  &  3  &  --  &   0.2  &    1.4  &  324.2239  &  -63.1143  &  1,2 \\
SPT-CLJ2137-6437  &  324.4178  &  -64.6235  &   4.60  &  --  &    0.91 $\pm$  0.07  &  3  &  --  &   3.7  &   12.8  &  324.4337  &  -64.6234  &  1 \\
SPT-CLJ2138-6007  &  324.5060  &  -60.1324  &  12.64  &  0.319  &    0.34 $\pm$  0.02  &  1  &  --  &  --  &  --  &  324.5036  &  -60.1317  &  2,4 \\
SPT-CLJ2139-5420  &  324.9669  &  -54.3396  &   4.81  &  --  &    0.24 $\pm$  0.02  &  1  &  --  &  --  &  --  &  324.9713  &  -54.3410  &  1 \\
SPT-CLJ2140-5331  &  325.0304  &  -53.5199  &   4.55  &  --  &    0.51 $\pm$  0.02  &  1  &  --  &  --  &  --  &  325.0287  &  -53.5037  &  1 \\
SPT-CLJ2140-5727  &  325.1380  &  -57.4564  &   5.08  &  --  &    0.40 $\pm$  0.03  &  1  &  --  &  --  &  --  &  --  &  --  &  2 \\
SPT-CLJ2142-4846  &  325.5693  &  -48.7743  &   4.53  &  --  &  $>$ 0.8  &  --  &  --  &  --  &   96.1  &  --  &  --  &  1 \\
SPT-CLJ2145-5644  &  326.4694  &  -56.7477  &  12.30  &  0.480  &    0.48 $\pm$  0.02  &  1  &  --  &  --  &  --  &  326.5298  &  -56.7422  &  2 \\
SPT-CLJ2146-4633  &  326.6473  &  -46.5505  &   9.59  &  0.933  &    0.95 $\pm$  0.07  &  3  &  --  &   0.2  &    1.5  &  --  &  --  &  1 \\
SPT-CLJ2146-4846  &  326.5346  &  -48.7774  &   5.88  &  0.623  &    0.62 $\pm$  0.02  &  2  &  --  &  --  &  --  &  326.5246  &  -48.7813  &  1 \\
SPT-CLJ2146-5736  &  326.6963  &  -57.6138  &   5.94  &  --  &    0.61 $\pm$  0.02  &  1  &  --  &  --  &  --  &  326.6954  &  -57.6310  &  2 \\
SPT-CLJ2148-4843  &  327.0971  &  -48.7287  &   4.64  &  --  &    0.98 $\pm$  0.07  &  3  &  --  &   0.6  &    1.1  &  --  &  --  &  1 \\
SPT-CLJ2148-6116  &  327.1798  &  -61.2791  &   7.27  &  0.571  &    0.52 $\pm$  0.02  &  1  &  --  &  --  &  --  &  327.1617  &  -61.2655  &  1,2 \\
SPT-CLJ2149-5330  &  327.3770  &  -53.5014  &   4.79  &  --  &    0.60 $\pm$  0.03  &  1  &  --  &  --  &  --  &  327.4331  &  -53.5176  &  1 \\
SPT-CLJ2150-6111  &  327.7177  &  -61.1954  &   4.70  &  --  &  $>$ 1.1  &  --  &  25.4  &  --  &   15.7  &  --  &  --  &  1 \\
SPT-CLJ2152-4629  &  328.1943  &  -46.4947  &   5.60  &  --  &  $>$ 1.5  &  --  &  20.0  &  10.6  &    8.0  &  --  &  --  &  1 \\
SPT-CLJ2152-5143  &  328.0034  &  -51.7245  &   4.53  &  --  &    0.41 $\pm$  0.03  &  1  &  --  &  --  &  --  &  327.9829  &  -51.7226  &  1 \\
SPT-CLJ2152-5633  &  328.1458  &  -56.5641  &   5.84  &  --  &  $>$ 1.5  &  --  &  --  &  20.2  &   55.5  &  --  &  --  &  1,2 \\
SPT-CLJ2155-5103  &  328.8747  &  -51.0508  &   4.52  &  --  &  $>$ 1.1  &  --  &  --  &  --  &   34.1  &  --  &  --  &  1 \\
SPT-CLJ2155-5225  &  328.8941  &  -52.4169  &   4.77  &  --  &    0.62 $\pm$  0.03  &  1  &  --  &  --  &  --  &  328.8997  &  -52.4194  &  1 \\
SPT-CLJ2155-6048  &  328.9850  &  -60.8072  &   5.24  &  0.539  &    0.48 $\pm$  0.02  &  1  &  --  &  --  &  --  &  328.9811  &  -60.8174  &  1 \\
SPT-CLJ2158-4702  &  329.6901  &  -47.0348  &   4.56  &  --  &  $>$ 0.9  &  --  &  --  &  --  &   64.4  &  --  &  --  &  1 \\
SPT-CLJ2158-4851  &  329.5737  &  -48.8536  &   4.61  &  --  &  $>$ 0.8  &  --  &  36.3  &  --  &   58.5  &  --  &  --  &  1 \\
SPT-CLJ2158-5615  &  329.5975  &  -56.2588  &   4.54  &  --  &  $>$ 1.1  &  --  &  --  &  --  &   53.6  &  --  &  --  &  1 \\
SPT-CLJ2158-6319  &  329.6390  &  -63.3175  &   4.54  &  --  &  $>$ 1.1  &  --  &  --  &  --  &   99.7  &  --  &  --  &  1 \\
SPT-CLJ2159-6244  &  329.9922  &  -62.7420  &   6.08  &  --  &    0.43 $\pm$  0.02  &  1  &  --  &  --  &  --  &  329.9944  &  -62.7539  &  2 \\
SPT-CLJ2200-5547  &  330.0304  &  -55.7954  &   4.80  &  --  &  $>$ 1.0  &  --  &  23.5  &  --  &   19.7  &  --  &  --  &  1 \\
SPT-CLJ2201-5956  &  330.4727  &  -59.9473  &  13.99  &  0.097  &    0.07 $\pm$  0.02  &  1  &  --  &  --  &  --  &  330.4723  &  -59.9454  &  2 \\
SPT-CLJ2202-5936  &  330.5483  &  -59.6021  &   4.89  &  --  &    0.42 $\pm$  0.03  &  1  &  --  &  --  &  --  &  330.5522  &  -59.6037  &  1 \\
SPT-CLJ2259-5432  &  344.9820  &  -54.5356  &   4.78  &  --  &    0.46 $\pm$  0.03  &  1  &  --  &  --  &  --  &  344.9765  &  -54.5260  &  3 \\
SPT-CLJ2259-5617  &  344.9974  &  -56.2877  &   5.29  &  --  &    0.15 $\pm$  0.02  &  1  &  --  &  --  &  --  &  345.0044  &  -56.2848  &  1,2 \\
SPT-CLJ2300-5331  &  345.1765  &  -53.5170  &   5.29  &  0.262  &    0.26 $\pm$  0.02  &  1  &  --  &  --  &  --  &  345.1655  &  -53.5199  &  2 \\
SPT-CLJ2301-5046  &  345.4585  &  -50.7823  &   4.58  &  --  &  $>$ 1.5  &  --  &  --  &  64.3  &   83.7  &  --  &  --  &  1 \\
SPT-CLJ2301-5546  &  345.4688  &  -55.7758  &   5.19  &  0.748  &    0.74 $\pm$  0.03  &  1  &  --  &   0.2  &    0.0  &  345.4595  &  -55.7842  &  1 \\
SPT-CLJ2302-5225 \tablenotemark{9}  &  345.6464  &  -52.4329  &   4.60  &  --  &  $>$ 1.0  &  --  &  --  &  --  &   83.3  &  --  &  --  &  1 \\
SPT-CLJ2311-5011  &  347.8427  &  -50.1838  &   4.64  &  --  &  $>$ 1.5  &  --  &  --  &  38.4  &   69.7  &  --  &  --  &  1 \\
SPT-CLJ2312-5820  &  348.0002  &  -58.3419  &   4.78  &  --  &    0.83 $\pm$  0.05  &  1  &  --  &   1.3  &    0.0  &  347.9912  &  -58.3428  &  1 \\
SPT-CLJ2329-5831  &  352.4760  &  -58.5238  &   4.95  &  --  &    0.81 $\pm$  0.03  &  1  &   2.3  &   0.0  &    0.1  &  352.4627  &  -58.5128  &  1 \\
SPT-CLJ2331-5051  &  352.9584  &  -50.8641  &   8.04  &  0.576  &    0.61 $\pm$  0.02  &  1  &  --  &  --  &  --  &  352.9631  &  -50.8650  &  2 \\
SPT-CLJ2332-5358  &  353.1040  &  -53.9733  &   7.30  &  0.402  &    0.38 $\pm$  0.02  &  1  &  --  &  --  &  --  &  353.1144  &  -53.9744  &  1,2 \\
SPT-CLJ2334-5953  &  353.6989  &  -59.8892  &   4.53  &  --  &  $>$ 1.5  &  --  &  --  &  71.7  &   26.1  &  --  &  --  &  1 \\
SPT-CLJ2337-5942  &  354.3544  &  -59.7052  &  14.94  &  0.775  &    0.76 $\pm$  0.03  &  1  &  --  &   0.3  &    0.0  &  354.3651  &  -59.7013  &  2 \\
SPT-CLJ2341-5119  &  355.2994  &  -51.3328  &   9.65  &  1.003  &    0.93 $\pm$  0.07  &  3  &  --  &   0.2  &    0.9  &  355.3015  &  -51.3290  &  1,2 \\
SPT-CLJ2342-5411  &  355.6903  &  -54.1887  &   6.18  &  1.075  &    0.96 $\pm$  0.07  &  3  &   7.4  &   2.4  &    9.6  &  355.6913  &  -54.1848  &  1$^{*}$ \\
SPT-CLJ2343-5521  &  355.7574  &  -55.3641  &   5.74  &  --  &  $>$ 1.5  &  --  &  --  &  66.3  &   50.8  &  --  &  --  &  1,2 \\
SPT-CLJ2343-5556  &  355.9290  &  -55.9371  &   4.58  &  --  &  $>$ 1.2  &  --  &  20.5  &  --  &    5.6  &  --  &  --  &  1$^{*}$ \\
SPT-CLJ2351-5452  &  357.8877  &  -54.8753  &   4.89  &  0.384  &    0.37 $\pm$  0.02  &  1  &  --  &  --  &  --  &  357.9086  &  -54.8816  &  1$^{*}$ \\
SPT-CLJ2355-5056  &  358.9551  &  -50.9367  &   5.89  &  0.320  &    0.28 $\pm$  0.02  &  1  &  --  &  --  &  --  &  358.9477  &  -50.9280  &  2 \\
SPT-CLJ2359-5009  &  359.9208  &  -50.1600  &   6.35  &  0.775  &    0.78 $\pm$  0.03  &  1  &  --  &   0.0  &    0.3  &  359.9284  &  -50.1672  &  2 \\

\enddata
\tablenotetext{a}{Spectroscopic redshift listed where available.  Details on references and observations are given in Table~\ref{tab:specz}.}
\tablenotetext{b}{Photometric redshift quality flag.  1-secure, 2-statistically inconsistent between three methods, 3-Swope or \spitzer\ IRAC colors only used, and  4-only one method used (except Swope and \spitzer-only case).  3$^*$ has larger bias correction (SWOPE only photometric redshift (see text for more detail).}
\tablenotetext{c}{Probability of finding a random position in the sky richer than the SPT cluster candidate, using single-band NIR galaxy overdensity.  Only calculated for unconfirmed candidates and confirmed clusters at $z>0.7$.}
\tablenotetext{d}{Cross-reference to imaging data.  Only the deepest imaging data source is noted in this table.  Internal references refer to Table~\ref{tab:oir cameras}. Ref. 1$^*$ indicates that BCS imaging data was used.}
\tablenotetext{1}{optical group on N z$\sim$0.3.}
\tablenotetext{2}{optical only z$\sim$1.0.}
\tablenotetext{3}{very complex region, optical group on NW z$\sim$0.4, another group on SW z$\sim0.65$.}
\tablenotetext{4}{optical group within 1' aperture z$\sim$0.35.}
\tablenotetext{5}{strong lensing arc.}
\tablenotetext{6}{strong lensing arc.}
\tablenotetext{7}{optical group on SW z$\sim$0.4.}
\tablenotetext{8}{optical group within 1' aperture z$\sim$0.15.}
\tablenotetext{9}{optical group on SE z$\sim$0.4.}

\end{deluxetable}
\end{center}
\clearpage

\end{landscape}

\pagestyle{plain}

\end{document}